\documentclass[a4paper,11pt]{JHEP3}
\usepackage[dvips]{graphicx}
\usepackage{amsmath}
\usepackage{epsfig}
\usepackage[english]{babel}
\usepackage[latin1]{inputenc}
\usepackage[T1]{fontenc}
\usepackage{amsfonts,amssymb}
\usepackage{float}
\usepackage{cite}
\usepackage{caption}

\newcommand{\vecL}{\left(\begin{array}{c}}
\newcommand{\vecR}{\end{array}\right)}

\title{Towards a  Systematic Construction of Realistic D-brane Models on a   del Pezzo Singularity}
\author{Matthew J. Dolan$^{1}$, Sven Krippendorf$^2$, Fernando Quevedo$^{2,3}$\\
$^{1}$ IPPP, Durham University, Durham DH1 3LE, UK\\
$^{2}$ DAMTP, University of Cambridge, Wilberforce Road, Cambridge, CB3 0WA, UK\\
$^{3}$ ICTP, Strada Costiera 11, Trieste 34151, Italy.
}

\abstract{A systematic approach is followed in order to identify realistic D-brane models at toric del Pezzo singularities. 
 Requiring quark and lepton spectrum and Yukawas from D3 branes and massless hypercharge, we are led to Pati-Salam extensions of the Standard Model.
Hierarchies of masses, flavour mixings and control of couplings select higher order del Pezzo singularities, minimising the Higgs sector prefers toric del Pezzos with $dP_3$ providing the most successful compromise. Then a supersymmetric local string model is presented with the following properties at low energies: (i) the MSSM spectrum plus a local $B-L$ gauge field or additional Higgs fields depending on the breaking pattern, (ii) a realistic hierarchy of quark and lepton masses and (iii) realistic flavour mixing between quark and lepton families with computable CKM and PMNS matrices, and CP violation consistent with observations.  In this construction, kinetic terms are diagonal and under calculational control suppressing standard FCNC contributions. Proton decay operators of dimension $4,5,6$ are suppressed, and gauge couplings can unify depending on the breaking scales from string scales at energies in the range $10^{12} - 10^{16}\, {\rm GeV},$ consistent with TeV soft-masses from moduli mediated supersymmetry breaking. The GUT scale model corresponds to $D3$ branes at $dP_3$  with two copies of the Pati-Salam gauge symmetry $SU(4)\times SU(2)_{R}\times SU(2)_{L}$ . $D-$brane instantons generate a non-vanishing $\mu-$term. Right handed sneutrinos can break the $B-L$ symmetry and induce a see-saw mechanism of neutrino masses and R-parity violating operators with observable low-energy implications. 
}
\preprint{DAMTP-2011-36\\
IPPP/11/32\\
DCPT/11/64}

\begin{document}
\newpage
\section{Introduction}
Constructing the Standard Model or a realistic extension within string theory is one of the biggest problems in string phenomenology. The challenge is  magnified due to two competing facts:  the large degeneracy of string models and the many experimentally verified properties of the Standard Model and evidence beyond. 

On the first issue, a large amount of discussion has been concentrated on the huge degeneracy of string vacua in four-dimensions, given by the number of Calabi-Yau compactifications, the choice of Standard Model embeddings, fluxes, etc. This lack of uniqueness complicates the  extraction of concrete model independent predictions of string theory that can be subject to experimental test (besides the standard `predictions' concerning the existence of gravity and other interactions,  dilaton and axion-like fields, moduli, extra spatial dimensions and supersymmetry but without setting the relevant mass scales).  Much discussion has been dedicated to this lack of uniqueness  leaving sometimes the  `existence' of realistic solutions almost as a non-issue.  

This second  issue is actually major since contrary to non-stringy model building, a successful string model has to be realistic in all aspects, that means it has to be consistent with all observational constraints in both high energy physics and cosmology. In particular 

\begin{itemize}

\item{}
It has to give rise to the gauge structure of the Standard Model with all the observed interactions, the three families of matter fields and at least one Higgs field or an  alternative mechanism of symmetry breaking.

\item{} It has to explain the hierarchy of masses of quarks and leptons, including neutrino masses.

\item{} It has to accommodate   all flavour issues such as the right mixing angles in the CKM and PMNS matrices and right amount of  CP-violation, preventing  the existence of unobserved FCNC.

\item{} It has to  explain the hierarchy of observed gauge couplings either by a consistent unification at the appropriate computable scale or otherwise.

\item{} The proton has to be stable enough to be consistent with observations while a concrete mechanism for baryogenesis has to be present.

\item{} It has to account for the dark matter of the universe. In particular it has to explain why extra matter fields do not cause cosmological problems, such as over-closing the universe.

\item{} It has to  account for inflation or other alternative early universe mechanism that addresses the same problems and gives rise to the density perturbations of the CMB. 

\item{} It has to address the hierarchy problem by low-energy supersymmetry or otherwise and has to determine dynamically all the relevant dimensional scales, such as the size and shape of all extra dimensions that determine the string and Kaluza-Klein scales in terms of the Planck mass, but also the electroweak scale. 

\item{} It has to have a mechanism to address the dark energy problem.

\end{itemize}

The important point that is usually overlooked is that even if a string model successfully accounts for most of the observable constraints, it takes  just one of these conditions  not to be satisfied to rule out the model. Due to this problem, much effort has been concentrated in identifying mechanisms in classes of models rather than in searching for explicit models. But this does not fully spare us from  the need to have explicit constructions of realistic models in order to be consistent with the standard claim that string theory is the best candidate for a unified theory of nature and to at least  serve as potential benchmark points for physics beyond the standard model. 

In the past decades there has been significant progress on building chiral string models of Particle Physics~\cite{ibanezuranga}. However to date there is no single compelling model that can incorporate the matter of the Standard Model, the spectrum of masses and the hierarchies in mixing among generations of quarks and leptons and the other points mentioned above. This is mainly due to lack of control over bulk geometries. This problem can be partially avoided by focusing on local model building. There, the bulk geometry can be locally controlled in the limit of decoupled gravity as for example in models built in intersecting D-brane models~\cite{0905.3379,0909.4292}, F-theory~\cite{0802.3391,0802.2969,0906.4672,1003.5337,1102.0290,0908.1784} or D-branes at singularities~\cite{0005067,0105042,0001083,Cascales:2003wn,0508089}, the latter of which we focus on in this article.
In such models the Standard Model matter content and couplings are completely determined by the local geometry and model building can be treated independently from other problems such as moduli stabilisation, dark energy,  or supersymmetry breaking. This approach is known as {\it bottom-up} model building, which was initiated  in~\cite{0005067}.

Local models are also perfectly suited to be incorporated in the success story of moduli stabilisation in type IIB string theory. This has allowed to determine all the relevant mass scales, addresses supersymmetry breaking and even the dark energy problem (although still subject to debate, it provides the mechanism that makes calculations neglecting this problem meaningful). Of particular interest is the so-called Large volume scenario \cite{0502058}, allowing for moduli stabilisation using $\alpha'$ and non-perturbative corrections, since it implements a concrete realisation of the modular approach to string model building. Typically, shrinking 4-cycles, the different 'modules', are responsible for the solutions of different physical problems such as moduli stabilisation, supersymmetry breaking, cosmological inflation \cite{0509012, 0808.0691}, and the realisation of the Standard Model. But at least one of the moduli is exponentially large and does not host the standard model, implying the standard model lives at branes on small cycles and then a local origin to the Standard Model. In this context we are interested in building supersymmetric models locally from the bottom-up. There are various ways of breaking supersymmetry in the local model, some of which construct a sector geometrically connected to the singularity as for instance in~\cite{0910.3616}, and others which rely on supersymmetry being broken by a geometrically separated sector such as in the Large volume scenario. The characteristic soft-masses for local models have been developed in \cite{0505076,0906.3297,1003.0388,1011.0999}.

In this article we focus on gauge theories arising from D-branes at toric singularities. This guarantees the absence of complex structure moduli in the superpotential. This class of models offers not only a very rich structure of interacting unified gauge theories with known superpotentials but also leading order control over the K\"ahler potential. Both the superpotential and K\"ahler potential are controlled by global symmetries, arising from the underlying geometric background. With this knowledge of the gauge theory at hand we want to demonstrate that these structures are sufficient for realistic Yukawa couplings both in quark and lepton sector. By restricting ourselves to model building in perturbative type IIB string theory the matter representations in combination with perturbative realisations of the Yukawa couplings restrict our model building to non simple-group grand-unified models (i.e.~we are restricted to models based not on $SU(5)$ or $SO(10)$ which is OK since this is not a requirement for coupling unification in these models). Then from the standard GUT models,  the largest unifying gauge groups we can utilise for model building are  the Pati-Salam  group, $SU(4)\times SU(2)_L \times SU(2)_R$~\cite{patisalam,Antoniadis:1988cm}, trinification $SU(3)^3$, etc. The Pati-Salam model is also the minimal model that realises hypercharge within the non-abelian gauge groups and has only one common Yukawa coupling for each family. From the singularities perspective we argue that the third del Pezzo singularity ($dP_3$) is the minimal one for which the hierarchy of Yukawas and flavour structure is achievable and it is the maximal that is also toric. These properties make $dP_3$ an optimal candidate to search for realistic models.  We will present the minimal model that not only includes the Standard Model matter content in the Pati-Salam gauge group but also realises its breakdown to the Standard Model spectrum. Further motivations for the Pati-Salam model are mentioned in the plethora of brane models based on Pati-Salam gauge groups~\cite{9805157,0202100,0403061,0510170,0708.2285,1002.2937,1104.2264}.

A crucial question after fixing the matter content is whether we can achieve the correct hierarchy in couplings for quarks and leptons. We obtain the correct quark flavour mixing as discussed in~\cite{1002.1790}. The difference in the lepton Yukawa couplings is achieved via a Majorana seesaw mechanism, which can be realised by giving a vev to a right-handed sneutrino. We show that the lepton flavour mixing can be different from the quark flavour structure. Using the right-handed neutrino masses obtained from giving a vev to the sneutrino, we can compute the PMNS neutrino mixing matrix. The precise values of the entries of the matrix depend on the details of the RH neutrino masses. However, our model unambiguously predicts that the mixing angle $\theta_{13}$ is greater than zero, in accord with recent results from the T2K experiment~\cite{Abe:2011sj}. The only toric model that can accommodate all criteria including all Standard Model matter is based on the third del Pezzo surface $dP_{3}.$

Gauge coupling unification occurs at the string scale $M_s$ or the winding scale $R M_s$ depending on the appearance of threshold effects~\cite{0901.4350,0906.1920}. The common gauge coupling at the unification scale, given by the dilaton, arises from the common geometric origin at the singularity for all different gauge factors therefore a simple group GUT is not required for unification. Depending on the breaking scale we find various possible unification scales ranging from an intermediate string scale at $M_s\sim 10^{12}\, {\rm GeV}$ to $M_s\sim 10^{16}\, {\rm GeV}.$ This flexibility in the unification scale is very attractive to allow for various string inspired models of supersymmetry breaking, partially requiring a string scale below the usual unification scale $M_s\sim 10^{16}{\rm GeV}.$ Various interesting low-energy phenomenological extensions to standard MSSM physics, including additional $U(1)$ symmetries at low energies, additional $SU(2)$ vector bosons or additional Higgses, are presented and can be roughly within reach for the LHC.

Proton decay operators are suppressed by $U(1)_{B-L}$ and anomalous $U(1)$ symmetries. If they are generated it is at the non-perturbative level after the breaking of $U(1)_{B-L}.$
We not currently have a dynamical mechanism to achieve the precise vevs responsible for breaking to the Standard Model or to minimise the scalar Higgs potential, but we do check for D- and F-flatness. A complete discussion of the latter requires more information about the embedding into the compact geometry and is hence tied to moduli stabilisation. 

The rest of this paper is organised as follows. We start with reviewing the gauge theory arising from del Pezzo singularities, argue for the special properties that make $dP_3$ appealing for model building and show how to embed the Pati-Salam gauge group in this setup. We then show how non-perturbative superpotential couplings induced from Euclidean 3-branes (E3) can give rise to the $\mu-$term. In Section 3 we discuss the breakdown of the Standard Model gauge groups, the masses of quarks and leptons and mixings. We also comment on the absence of proton decay and analyse gauge coupling unification in this model and finally in Section 4 we discuss whether the breaking scales can be combined with $TeV-$scale soft-masses.

%%%%%%%%%%%%%%%%%%%%%%%%%%%%%%%%%%%%%%%%%%%%%%%%%%
\section{A Review of the Ingredients}
%%%%%%%%%%%%%%%%%%%%%%%%%%%%%%%%%%%%%%%%%%%%%%%%%%
\subsection{Models at del Pezzo singularities}
There is an infinite number of gauge theories from branes at singularities. The largest class of singularities where we control the gauge theory are based on toric singularities. 

The simplest examples of such singularities are the del Pezzo surfaces $dP_0-dP_3$ where one cycle collapses to zero size and which, fortunately, allow for interesting low-energy physics. Recall that the $n^{th}$ del Pezzo surface $dP_n$ $(n=0,\ldots,8)$ corresponds to $\mathbb{P}^2$ blown-up at up to eight points, the first four $dP_0-dP_3$ being toric. The gauge theories probing these singularities can be obtained using dimer techniques (for a review of the del Pezzo examples see for example the discussion in \cite{1002.1790}). Up to $dP_4$ no complex structure moduli appear in the superpotential~\cite{0212021}, the value of complex structure moduli for higher del Pezzos depends on the moduli stabilisation scenario, hence rendering the analysis of couplings more difficult and less attractive. In addition $dP_3$ is sufficient for our purposes, and allowing us to stop there. 

Historically local model building started on $\mathbb{C}^3/\mathbb{Z}^3=dP_0$ \cite{0005067}. Although a realistic Standard Model-like matter content can be achieved, the mass eigenvalues are found to be of the form $(0,M,M)$, rendering the model un-realistic. To change the couplings one can either consider non-trivial B-flux  threading the cycles of the singularity~\cite{0512122}, which we shall not pursue further in this article, or consider singularities which allow for richer couplings {\it per se}. In~\cite{0810.5660} it was realised that  models based on $dP_1$ lead to a potentially hierarchical mass structure of the form $(0,m,M).$ As discussed in~\cite{1002.1790}, the structure of couplings is in-sufficient to achieve the hierarchical mixing angles in the CKM matrix, favouring models based on $dP_2$ or higher del Pezzos. As we are interested in models where all Standard Model fields arise from $D3-D3$ states, we need a mechanism to discriminate between couplings for quarks and leptons. As discussed later in Section~\ref{sec:mixing}, this requires an extended field content requiring to consider models based on $dP_3.$\footnote{It would be very interesting to see whether this constraint can rule out models based on lower del Pezzos whose coupling structure is changed with B-fluxes. Since this option 'only' allows for changing the coefficient in front of every coupling, this option does only change the Pati-Salam Yukawa coupling but not Standard Model Yukawa couplings for quarks and leptons separately.} Hence from now on we focus on models based on the $dP_3$ singularity.

\subsection{Models from the $dP_3$ singularity}
The third del Pezzo surface $dP_3$ features six $U(N)$ gauge groups and there are four toric phases connected via Seiberg duality~\cite{0205144}. Here we choose the toric phase that is connected simply via Higgsing to the lower del Pezzo surfaces, its matter content is shown in Figure~\ref{fig:dp3general}. Recall that every node corresponds to an $U(N)$ gauge group and an arrow from node $A$ to $B$ in the quiver corresponds to a field $X_{AB}$ transforming in the bi-fundamental $(N_A,\bar{N}_B)$ under $U(N_A)\times U(N_B).$ The superpotential is given by
\begin{eqnarray}
\nonumber W_{dP_{3}}&=&-X_{12} Y_{31} Z_{23}-X_{45} Y_{64} Z_{56}+X_{45} Y_{31} Z_{14} \frac{\rho _{53}}{\Lambda}+X_{12} Y_{25} Z_{56} \frac{\Phi _{61}}{\Lambda}\\ \nonumber & &+X_{36} Y_{64} Z_{23} \frac{\Psi _{42}}{\Lambda}-X_{36}Y_{25} Z_{14} \frac{\rho _{53} \Phi _{61} \Psi _{42}}{\Lambda^3}\\ \vspace{14pt}
&=&\left(
\begin{array}{c}
 X_{45} \\
 Y_{25} \\
 Z_{23}
\end{array}
\right)\left(
\begin{array}{ccc}
 0 & Z_{14} \frac{\rho _{53}}{\Lambda} & -Y_{64} \\
 -Z_{14} \frac{\rho _{53} \Phi _{61} \Psi _{42}}{\Lambda^3} & 0 & X_{12} \frac{\Phi _{61}}{\Lambda} \\
 Y_{64} \frac{\Psi _{42}}{\Lambda} & -X_{12} & 0
\end{array}
\right)\left(
\begin{array}{c}
 X_{36} \\
 Y_{31} \\
 Z_{56}
\end{array}
\right),
\label{eq:dp3superpotential}
\end{eqnarray}
where $\Lambda$ is an appropriate UV cutoff.\footnote{Following~\cite{0611144} we assume that within the low energy $\mathcal{N}=1$ supergravity,
$\Lambda$ is $M_P$ due to holomorphy, and the actual physical suppression scale is determined by terms in the K\"ahler potential.}
\begin{center}
\includegraphics[width=0.5\textwidth]{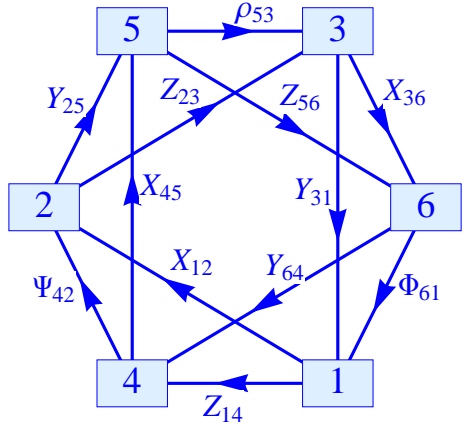}
\captionof{figure}{\footnotesize{The $dP_3$ quiver with fields labelled in correspondence to the superpotential in Equation~\ref{eq:dp3superpotential}.}\label{fig:dp3general}}
\end{center}
The structure of the superpotential is entirely fixed by the geometry, which - from a field theoretical point of view - can be seen as a global $E_3=SU(2)\times SU(3)$ symmetry and an additional R-symmetry \cite{0404065}. The charges of the fields under the global symmetries are summarised in Table~\ref{tab:globalsymmetries}. The superpotential can be determined from invariance under these symmetries.
\begin{center}
\begin{tabular}{c | c | c}
Fields & $SU(2)\times SU(3)$ & $U(1)_R$ \\ \hline\hline
$(X_{36}^R,\,Y_{25}^L,\,Z_{14},\,\Psi_{42},\,\rho_{53},\,\Phi_{61})$ & $(2,3)$ & 1/3\\
$(X_{45}^L,\, Z_{56}^R,\,Y_{64})$ & $(1,\bar{3})$ & 2/3\\
$(Z_{23}^L,\, Y_{31}^R,\,X_{12})$ & $(1,\bar{3})$ & 2/3\\
\end{tabular}
\captionof{table}{\footnotesize{Charges of $D3-D3$ states under global symmetries, taken from \cite{0404065}. The last two rows are written separately, since gauge invariance forbids a coupling between them.}\label{tab:globalsymmetries}}
\end{center}

Choosing two copies of the Pati-Salam gauge group as shown in Figure~\ref{fig:psdp3breaking}, we have that all Standard Model fields are given by $D3-D3$ states. 
\begin{center}
\includegraphics[width=0.5\textwidth]{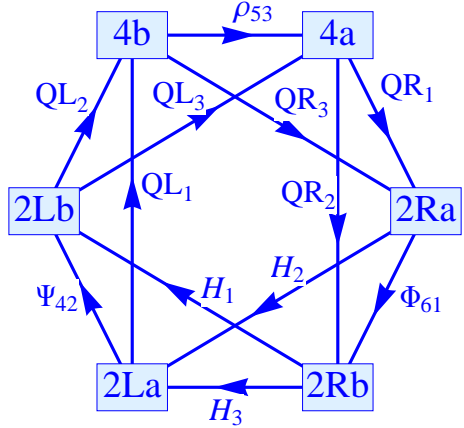}
\captionof{figure}{\footnotesize{The quiver diagram of the Pati-Salam realisation on $dP_3.$ The fields $(\Phi_{61},\Psi_{42},\rho_{53})$ will be responsible for gauge group breaking, combining the two copies of every gauge group. The other fields are Standard Model fields including three generation of Higgs fields.
}\label{fig:psdp3breaking}}
\end{center}
In particular we identify the fields $Q_{L}^{i}=(X_{45},Y_{25},Z_{23})$ as left-handed quarks (and leptons), $Q_{R}^{i}=(X_{36}^R,Y_{31}^R,Z_{56}^R)$ as right-handed quarks (and leptons) and the matrix as the Yukawa matrix. The fields $H_{i}=(X_{12},Y_{64},Z_{14})$ are Higgs fields. With this identification we can re-write the superpotential in the usual form
\begin{eqnarray}
\nonumber W_{\rm D3D3}&=&-H_{1} Q_{2}^R Q_{3}^L-Q_{1}^L H_{2} Q_{3}^R+Q_{1}^L Q_{2}^R H_{3} \frac{\rho _{53}}{\Lambda}+H_{1} Q_{2}^L Q_{3}^R \frac{\Phi _{61}}{\Lambda}\\ \nonumber & &+Q_{1}^R H_{2} Q_{3}^L \frac{\Psi _{42}}{\Lambda}-Q_{1}^RQ_{2}^L H_{3} \frac{\rho _{53} \Phi _{61} \Psi _{42}}{\Lambda^3}\\ \vspace{14pt}
&=&\left(
\begin{array}{c}
 Q_{1}^L \\
 Q_{2}^L \\
 Q_{3}^L
\end{array}
\right)\left(
\begin{array}{ccc}
 0 & H_{3} \frac{\rho _{53}}{\Lambda} & -H_{2} \\
 -H_{3} \frac{\rho _{53} \Phi _{61} \Psi _{42}}{\Lambda^3} & 0 & H_{1} \frac{\Phi _{61}}{\Lambda} \\
 H_{2} \frac{\Psi _{42}}{\Lambda} & -H_{1} & 0
\end{array}
\right)\left(
\begin{array}{c}
 Q_{1}^R \\
 Q_{2}^R \\
 Q_{3}^R
\end{array}
\right),
\end{eqnarray}
which we shall use from now on.

\subsection{K\"ahler potential}
In a purely local model the global symmetries or isometries restrict the K\"ahler potential to a diagonal form. However this is no longer guaranteed in an effective supergravity setup where these global symmetries are broken by volume dependent effects. The precise scale at which they are broken is not known since control over the K\"ahler potential is very limited. However, the extended gauge structure of $dP_3$ provides us with new control over the appearance of the matter fields in the K\"ahler potential.\footnote{Similar use of multiple $U(1)$ symmetries has also been made in recent heterotic model building~\cite{1106.4804}.}

Since there are only single fields between any two gauge groups, the constraint of gauge invariance allows only flavour diagonal terms in the K\"ahler potential at leading order. Since all Standard Model matter fields have the same geometric origin, they will have a common overall factor depending on the K\"ahler moduli. To leading order in the $1/{\cal V}$ large volume expansion, we can write
\begin{equation}
K_{\rm matter}\supset \frac{a+f(\tau_s,\tau_b)}{{\cal V}^{2/3}}\left(Q^i_{L,R} \bar{Q}^{i}_{L,R}+H_{i}\bar{H}_{i}+\Phi_{61}\bar{\Phi}_{61}+\Psi_{42}\bar{\Psi}_{42}+\rho_{53}\bar{\rho}_{53}\right),
\end{equation}
where $f(\tau_s,\tau_b)$ is a function of the small modulus being suppressed by higher inverse powers in the volume.\footnote{The precise structure of this modulus weight is not known but can be estimated in various setups \cite{0609180, 0805.2943}. The structure of the moduli weights is of particular importance for the soft masses as discussed in \cite{0906.3297, 0912.2950}, and  in the context of guaranteeing F-flatness.} This structure also ensures that the soft-masses induced via moduli mediation in a large volume setup are flavour-diagonal, satisfying the conditions presented in~\cite{0710.0873}. In particular, problems for flavour changing neutral currents arising from a scalar mass matrix that is not proportional to the K\"ahler metric $K_{ab}$ are absent as long as the gauge symmetries are unbroken~\cite{0912.2950}.
We also note that terms like $H_{3}H_{3}+\bar{H}_{3}\bar{H}_{3},$ utilised in the Giudice-Masiero mechanism, are forbidden by the anomalous $U(1)$ symmetries.

%%%%%%%%%%%%%%%%%%%%%%%%%%%%%%%%
\subsection{Anomaly cancellation and $D7$ branes}
%%%%%%%%%%%%%%%%%%%%%%%%%%%%%%%%
In order to cancel anomalies for the given choice of $D3$ brane gauge groups, it is necessary to have $D7$ branes that lead to additional $D7-D3$ states. The spectrum of $D7-D3$ states for general toric singularities was developed in \cite{0604136}, leading to one pair of $D3-D7$ and $D7-D3$ states for every $D3-D3$ state. For every $33$ state $\Phi_{3_i 3_j}$, there exists a supersymmetric 7-brane giving a $(7i)$ fundamental and a
$(7j)$ anti-fundamental with the Yukawa coupling $\Phi_{3_i 3_j} (7i) (7j)$. The most general $dP_3$ quiver including $D7$ branes is shown in Figure~\ref{fig:dp3d7quiver}. As discussed for example in \cite{0803.4474}, the cancellation of anomalies is checked by counting the arrows coming in/out to any node weighted by the rank of the gauge group they originate/end. In our (Pati-Salam$)^2$ model we find the following constraints for the $D7$ rank gauge groups:
\begin{eqnarray}
\nonumber m_{12}&=& 4 - m_4 + m_5 + m_6\, ,\\
\nonumber m_{11}&=& 2 - m_1 + m_2 + m_4\, ,\\
\nonumber m_{10}&=&4 - m_1 + m_3 + m_7\, ,\\
\nonumber m_{9}&=& m_2 - m_3 + m_5 + m_6 - m_7\, ,\\
m_{8}&=& -2 + m_2-m_3+m_5\, .
\label{anomalycanc}
\end{eqnarray}
Solutions with $m_{i} \geq 0$
are physically relevant. Note that although we have in principle six anomaly cancellation conditions, one of them is trivially satisfied when the other five are satisfied. In section \ref{sec:modelatlowenergies} (cf. Figure \ref{d3d3d3d37}) we specify a  choice for the $D7$ gauge groups compatible with the breakdown to the Standard Model gauge symmetries. We note that the additional $D7-D3$ states can decouple from the low-energy effective action, since $D7-D7$ interactions in the bulk away from the local construction can give rise to large masses.

\begin{center}
\includegraphics[width=0.75\textwidth]{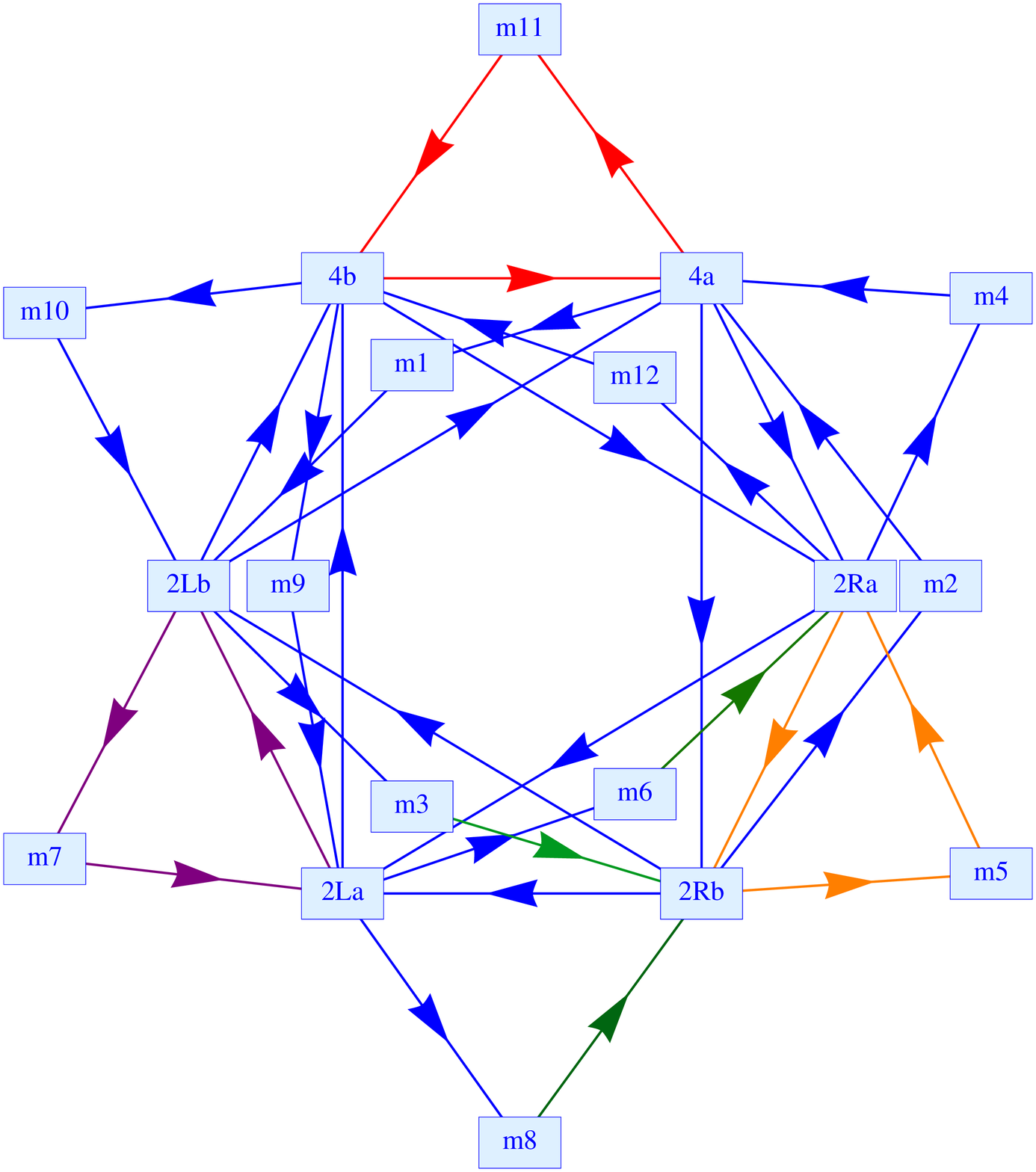}
\captionof{figure}{\footnotesize{The $dP_3$ quiver with $D7$ branes included. Indicated in red are the fields that we vev to break the $U(4)\times U(4)$ symmetry, the $D3-D7$ states need vevving to satisfy D-term equation. Similarly the states which break the left and right $U(2)$ symmetries are indicated in purple and orange. The states highlighted in green are $D7$ states that we could vev in accordance with the D-term equations to give large masses to some of the Higgs fields.}\label{fig:dp3d7quiver}}
\end{center}
%%%%%%%%%%%%%%%%%%%%%%%%%%%%%%%%%%%%%%%%%%%
\subsection{Anomalous $U(1)$s}
\label{sec:anomu1}
%%%%%%%%%%%%%%%%%%%%%%%%%%%%%%%%%%%%%%%%%%%
The $U(1)$ factors contained in $U(N)$ factors appear differently in the low-energy description depending on their origin. We distinguish between:
\begin{enumerate}
\item $U(1)$ symmetries within non-abelian symmetries such as the diagonal generators in $SU(N).$ These $U(1)$ symmetries remain massless at low-energies.
\item Anomalous $U(1)$ symmetries get string scale masses.
\item Non-anomalous $U(1)$ symmetries remain massless in the non-compact model.  However, once embedded into a compact model they remain only massless if the associated cycle becomes trivial in the bulk,~\cite{0610007}. Otherwise they receive masses via the Green-Schwarz mechanism.
\end{enumerate}
As shown in \cite{0810.5660} the masses for non-anomalous $U(1)s$ upon compactification are suppressed compared to the masses for anomalous $U(1)$ symmetries and are given by the mass for KK-modes $M_{KK}\sim 1/{\cal V}^{2/3}.$
In the philosophy of bottom-up model building we demand that all cycles associated to non-anomalous $U(1)$ symmetries are non-trivial in the bulk. We then are left with only non-abelian $SU(N)$ symmetries. The now massive $U(1)$ symmetries remain as global symmetries in the low-energy spectrum.

%%%%%%%%%%%%%%%%%%%%%%%%%%%%%%%%%%%%%%%%%%%%
\subsection{E3 branes and non-perturbative effects}
\label{sec:nonpert}
%%%%%%%%%%%%%%%%%%%%%%%%%%%%%%%%%%%%%%%%%%%%
Non-perturbative contributions to the superpotential can be induced from Euclidean $E3$ branes wrapped on 4-cycles passing through the singularities in analogy to $D7$ branes.  Instanton induced superpotentials for branes at singularities with $E3$ branes have been studied in \cite{0711.1316} (for a review of the subject see~\cite{0902.3251}). The reader is referred to this reference for a derivation of the conditions required for a sufficient zero mode structure. We will utilise these effects to generate $\mu-$terms for the Higgs fields.

For our purpose the additional zero modes between $E3$ branes and $D3/D7$ branes are of interest. The spectrum and couplings of these zero modes are the same as if there were a $D7$ brane wrapping the E3-wrapped cycle.

The presence of the desired couplings however does not imply that a non-perturbative contribution to the superpotential is generated, one has to ensure the correct number of fermion zero modes. The classical action for the instanton is given by a global piece depending on the 4-cycle volume they are wrapping, a local piece depending on the twisted moduli at the singularity and a piece coming from couplings to bi-fundamental fields
\begin{equation}
S_{E3^r}=S_{E3^r}^{\rm unt}+S_{E3^r}^{\rm twisted}=T^r+\sum d^r_k \phi_k +\sum_{i,j,r} c_{ij}^r \eta_i^\alpha \Phi_r^{\alpha\beta}\eta_j^\beta\, ,
\end{equation}
where $\eta_x$ denotes the corresponding zero mode with the $E3$ and $D3$ brane, $\phi_k$ are twisted closed string moduli which we can neglect from now on, $\Phi_r$ denotes a $33$ state, and  $c_{ij}$ is a coupling constant, for our purposes mainly indicating whether the coupling exists or not. In addition we have couplings among $D3-D7,$ $E3-D3,$ and $D7-E3$ zero modes of the general form
\begin{equation}
\eta_{E3D7}\Phi_{37}\eta_{E3D3}\, .
\label{e3d7coupling}
\end{equation}
There can be further zero modes involving $D7-D7$ states depending on the boundary conditions at infinity which we shall not need in the further discussion. We need to integrate over the fermionic zero modes which gives rise to the following non-perturbative superpotential
\begin{equation}
e^{-S_{E3^r}}\int [d\eta^\alpha][d\bar{\eta}^\beta][d\tilde{\eta}^\gamma]\, e^{-\sum_{i,j,r} c_{ij}^r \eta_i^\alpha \Phi_r^{\alpha\beta}\bar{\eta}_j^\beta-\sum_{l,m}\tilde{c}^s_{l,m}\tilde{\eta}^{\gamma}_l\Phi_{37_s}^{\gamma\alpha}\eta^{\alpha}_m}\, ,
\label{zeromodes}
\end{equation}
where $\tilde{\eta}^\gamma$ refers to the E3-D7 zero modes. In order to give a non-vanishing contribution every zero mode has to appear precisely once. The scenario presented below for the $\mu-$term should be seen as to show how such a contribution can arise, keeping in mind that depending on the boundary conditions at infinity there could be more zero modes, which can change the contribution by additional vevs appearing in the $\mu-$term.

%%%%%%%%%%%%%%%%%%%%%%%%%%%%%%%%%%%%%%%%%%%%
\subsubsection{ The $\mu$-term}
%%%%%%%%%%%%%%%%%%%%%%%%%%%%%%%%%%%%%%%%%%%%
For a viable low-energy Higgs sector we need to have a $\mu-$term for the Higgs fields $\mu H_u.H_d.$ In a model with left-right extension the Higgs field $H$ including both $H_u$ and $H_d$ fields, the $\mu-$term is conveniently rewritten as $\mu H.H$ where the product $H.H$ is to be understood as a contraction of the $SU(2)$ indices with $\varepsilon_{\alpha\beta}.$

In models from branes at singularities the Higgs fields are additionally charged under anomalous $U(1)$ symmetries which forbid this coupling perturbatively. We remind the reader here about the arguments presented in \cite{0711.1316} on how to generate this coupling non-perturbatively via stringy instantons from E3-branes. To project down to the required number of two fermionic zero modes we have to require that either  the $E3$ brane is mapped to itself by an orientifold projection away from the singularity and that the singularity is mapped to a mirror image of itself or alternatively that a single $D7$ brane is wrapped on that cycle \cite{0711.1837}. In the bottom-up philosophy we are not looking for a concrete geometric realisation of the compact setup, but we require that anomalies are cancelled locally.

The Higgs field $H$ transforms as $(2_R, \bar{2}_L)$ and there are the following charged zero modes in this scenario between the $D3$ and $E3$ brane:
\begin{equation}
\eta^L=(1,2_L,1)\hspace{1cm}\text{ and } \eta^{R}=(\bar{2}_R,1,1)\, ,
\end{equation}
where the last charge denotes the charge under the $E3$ brane gauge group.

\noindent In this setup the following couplings are induced
\begin{equation}
S_{\rm charged}=\eta^L H \eta^R\, .
\end{equation}
Now we can integrate over the charged zero modes as follows
\begin{eqnarray}
S_{\rm charged}&=&\int d\eta^L_{1,2}d\eta^R_{1,2}\, \eta^L_{i} H_{ij} \eta^R_j\\
&=& \varepsilon^{ac}\varepsilon^{\dot{b}\dot{d}}H_{a\dot{b}}H_{c\dot{d}}\, .
\end{eqnarray}
This then leads to the following non-perturbative contribution to the superpotential as outlined above
\begin{equation}
W_{\rm np}=A e^{-a T_s} H.H\, ,
\end{equation}
where $T_s$ denotes the chiral superfield associated to the 4-cycle volume which the E3-brane is wrapping and $A$ denotes a constant depending on complex structure moduli. Note that this allows us to obtain different vevs for the Higgs fields due to different cycles the E3-branes are wrapping. Note that we do not have any mixing among Higgs field generations through the $\mu$ term in the superpotential.

%%%%%%%%%%%%%%%%%%%%%%%%%%%%%%%%%%%%%%%%%
\subsection{Neutrino see-saw mechanism}
%%%%%%%%%%%%%%%%%%%%%%%%%%%%%%%%%%%%%%%%%
At tree-level there is no Majorana mass term due to the additional $U(1)$ symmetries of the singularity. There are two mechanisms which could generate Majorana neutrino masses:
\begin{enumerate}
\item A vev for the right-handed sneutrino can give rise to Majorana neutrino masses as for instance in~\cite{0710.3525,0904.4509,0910.1129,1005.5392,0811.3424,0812.3661,Anderson:2011ns}. We postpone the discussion of this mechanism to section~\ref{sec:rparity} since it depends on the structure of vevs we demand for a viable flavour structure.

\item D-brane instantons could lead to the desired Majorana masses.

\end{enumerate}

In Appendix~\ref{sec:NPneutrinos} we show that the second possibility cannot be realised in the model presented in this paper. The neutrino mass term is forbidden by the $U(1)_{B-L}$ gauge symmetry, which will not be broken by non-perturbative effects. However, if we were to break this symmetry by giving a vev to a 37 string state, the symmetry would become an effective global symmetry which could be broken by the string instantons. After doing this, both the $U(1)_{B-L}$ and $U(1)_Y$ groups mix with the abelian $U(1)$ factors from the D7 gauge groups. This in turn implies that the K\"ahler modulus is charged under these symmetries, and generates a St\"uckelberg mass for both groups. There is therefore a tension between having a non-perturbatively generated mass for the right-handed neutrinos and keeping hypercharge unbroken. By extending the D7 sector we have found it possible to generate a mass term for the neutrinos; however such a term comes from a higher dimensional operator leading to un-physically low neutrino masses. Of course, it still may be possible to have viable instantonic neutrino masses in models based on different gauge groups, or at different, possibly non-toric singularities. The construction of models at toric singularities with non-perturbative neutrino masses is an interesting challenge.

%%%%%%%%%%%%%%%%%%%%%%%%%%%%%%%%%%%%%%%%%%%%%%
\section{The Standard Model in $dP_3$}
%%%%%%%%%%%%%%%%%%%%%%%%%%%%%%%%%%%%%%%%%%%%%%

After discussing all the ingredients of our model, let us now focus on how the Standard Model matter content and couplings can be accommodated in the third del Pezzo surface $dP_3.$ We start by discussing how the gauge structure in $dP_3$ can be broken to the Standard Model gauge group. We specify the $D7$ sector and determine the matter content below the breaking scale. After that we discuss the phenomenology in terms of the masses for quarks and leptons, the flavour mixing, proton decay, the number of parameters we tune and gauge coupling unification.

%%%%%%%%%%%%%%%%%%%%%%%%%%%%%%%%%%%%%%%%%%%%%%%
\subsection{Breakdown to the Standard Model}
%%%%%%%%%%%%%%%%%%%%%%%%%%%%%%%%%%%%%%%%%%%%%%%

The breaking of gauge symmetries can be achieved by resolving the singularity slightly, which corresponds in the gauge theory to non-vanishing FI-terms. To satisfy the D-term equations then requires certain fields to acquire vevs. Here we only check that our vev assignments are actually flat directions of the potential without completely minimising the potential. This is due to the lack of knowledge of how to stabilise the moduli associated to the slight resolution of the singularity. In a complete model, the 4-cycle volume associated to the Standard Model singularity corresponds to a modulus and needs to be stabilised as well. In local D-brane models at del-Pezzo singularities the singular point corresponds to a zero 4-cycle volume. For the del Pezzo 4-cycle the usual non-perturbative effects cannot be used for stabilisation \cite{0711.3389}, the leading contribution then arises from D-term potentials associated to anomalous $U(1)$ symmetries which the 4-cycle volume is charged under. Having plenty of anomalous $U(1)$ symmetries associated with the singularity it is natural to assume that the del Pezzo 4-cycle modulus will be charged under them. At tree-level, this implies a stabilisation in the singular regime, explicitly driving the 4-cycle volume to zero size. Higher order effects then will become important to determine the size of this breaking.

A detailed analysis of the stabilisation is beyond the scope of this paper and we only pursue the search for interesting flat directions at this stage.

From now we allow ourselves to set the scale of breaking down to the MSSM to be any scale below the string scale. Whether we can break directly to the Standard Model or in an intermediate step to a left-right model is dealt with in section~\ref{sec:gaugecoupling}.

To break $SU(4)$ we use a method that appeared in the context of $SU(5)$ GUT models \cite{9404278}, starting with two copies of $SU(4)$ and breaking it to diagonal gauge groups with bi-fundamental matter transforming as $(4,\bar{4}).$ Our approach differs from \cite{0202100} through the presence of only one bi-fundamental field between the $SU(4)$ factors, possible through the appropriate inclusion of $D3-D7$ states. 
In the $dP_3$ model, we have precisely one field $\rho_{53}$ transforming as $(4,\bar{4}).$ Giving $\rho_{53}$ a vev allows for the following breakdown:
\begin{eqnarray}
\langle \rho_{53}\rangle_{2}&=&\left(\begin{array}{c c c c}
v_1 & 0 & 0 & 0\\
0 & v_1 & 0 & 0\\
0 & 0 & v_1 & 0\\
0 & 0 & 0 & v_2
\end{array}\right):\; U(4)_a \times U(4)_b \to SU(3)_{\rm diag}\times U(1)_4 \times U(1)_d\, ,
\label{eq:breaking1}
\end{eqnarray}
where $U(1)_4$ refers to the $U(1)$ in $SU(4)$ and $U(1)_d$ is the diagonal $U(1)$ of the two $U(1)$ factors from $U(4).$ 
 The D-flat conditions for this vev can be guaranteed by the presence of $D3-D7$ states with appropriate vev as already mentioned in Equation~\ref{rho7vev}. The two $U(2)_L$ gauge groups are broken to the diagonal subgroup by vevving $\Psi_{42}$ as follows:
\begin{equation}
\langle\Psi_{42}\rangle=\left(\begin{array}{c c}\psi & 0\\ 0 &\psi\end{array}\right).
\label{eq:breaking2}
\end{equation}
This vev satisfies the $SU(2)$ D-term equations by itself. The $U(1)$ equations can be satisfied by the appropriate FI-term.
The two $U(2)_R$ factors can either be broken to the diagonal gauge group or directly down to $U(1)\times U(1)$ via the following vev
\begin{equation}
\langle\Phi_{61}\rangle=\left(\begin{array}{c c}\phi & 0\\ 0 &\tilde{\phi}\end{array}\right).
\label{eq:breaking3}
\end{equation}
As in the case for the breaking of the U(4) factors, the D-flatness can be guaranteed by the presence of $D3-D7$ states as in Equation~\ref{phi7vev}.
Further details on the breaking including a discussion of the masses associated to these breaking fields and how sufficient F-flatness is achieved can be found in Appendix~\ref{sec:consistency}.

%%%%%%%%%%%%%%%%%%%%%%%%%%%%%%%%%%%%%%%%%%%%%%
\subsection{The model at low-energies}
\label{sec:modelatlowenergies}
%%%%%%%%%%%%%%%%%%%%%%%%%%%%%%%%%%%%%%%%%%%%%%
The presence of appropriate D-instanton effects as discussed in previous sections fixes our choice of $D7$ gauge groups almost entirely. The only flexible choices are $m_6$ and $0\leq m_9\leq 2.$ The most general choice of $D7$ gauge groups in this context is shown in figure~\ref{d3d3d3d37}. We choose the ranks to satisfy the anomaly cancellation condition (\ref{anomalycanc}) and still to allow for consistent breakdown of the Pati-Salam gauge group to the Standard Model.\footnote{This restriction is not possible when considering neutrino Majorana masses induced via a vev for the sneutrino as discussed in section~\ref{sec:rparity}. A model without $D7$ branes needed for anomaly cancellation is presented in Appendix~\ref{sec:nod7s}.}
\begin{center}
\includegraphics[width=0.7\textwidth]{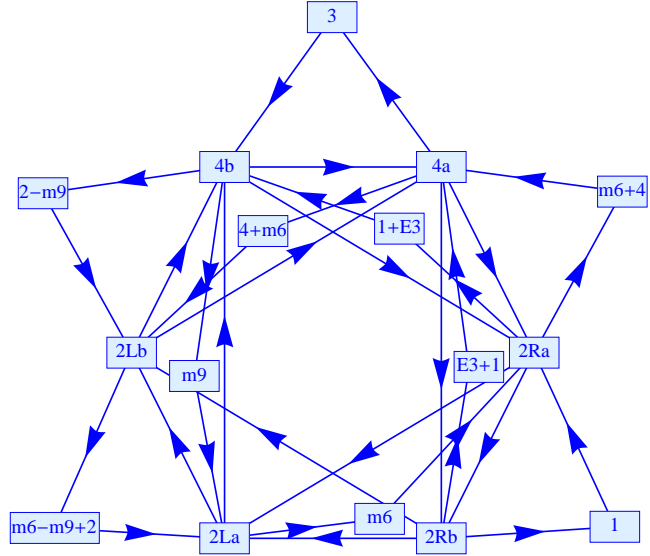}
\captionof{figure}{\label{d3d3d3d37}\footnotesize{The $D3-D3$ and $D3-D7$ spectrum in the Pati-Salam model. The ranks for the $D7$ branes are chosen to satisfy the anomaly constraints and allow for breaking to the Standard Model gauge group.}}
\end{center}
As discussed in section~\ref{sec:anomu1}, we assume that all abelian $U(1)$ factors of the $D3$ gauge groups become massive. We identify the non-anomalous linear combination of $T^3_{\rm SU(2)_R}$ and $U(1)_4,$ the diagonal $U(1)$ in $SU(4),$ as hypercharge. 

The vevs in equations \ref{eq:breaking1}, \ref{eq:breaking2} and \ref{eq:breaking3} lead to the $D3-D3$ spectrum after symmetry breaking shown in Table~\ref{d3d3spectrum}.
\begin{center}
\begin{tabular}{ c | c | c c c | c}
total \# & Fields & $SU(3)$ & $SU(2)$ & $U(1)_Y$ & $U(1)_x$\\ \hline
$3$ & $Q^L_{1},$ $Q^L_{2},$ $Q^L_{3}$ & 3 & $\bar{2}$ & $a$  & $a$ \\
$3$ & $u_{1},$ $u_{2},$ $u_{3}$ & $\bar{3}$ & 1 & $-a+k$  & $-a-k$ \\
$3$ & $d_{1},$ $d_{2},$ $d_{3}$ & $\bar{3}$ & 1 & $-a-k$& $-a+k$ \\ \hline
$3$ & $L_{1},$ $L_{2},$ $L_{3}$ & 1 & $\bar{2}$ & $-3a$& $-3a$ \\
$3$ & $\nu_{1},$ $\nu_{2},$ $\nu_{3}$ & $1$ & 1 & $3a+k$& $3a-k$ \\
$3$ & $e_{1},$ $e_{2},$ $e_{3}$ & $1$ & 1 & $3a-k$& $3a+k$\\ \hline
$3$ & $H^u_{1},$ $H^u_{2},$ $H^u_{3}$ & 1 & $2$ & $-k$ & $k$\\
$3$ & $H^d_{1},$ $H^d_{2},$ $H^d_{3}$ & 1 & $2$ & $k$ & $-k$\\
\end{tabular}
\captionof{table}{\footnotesize{The $D3-D3$ spectrum after symmetry breaking. For correct hypercharge assignments we need in our conventions $a=-1/6$ and $k=1/2.$}\label{d3d3spectrum}}
\end{center}
$U(1)_x$ is generated by $T_{15}^{SU(4)}-T_{3}^{SU(2)_R}.$ This additional $U(1)$ symmetry, which is just the difference between $B-L$ and hypercharge, can be broken dynamically below the string scale for instance utilising the right-handed sneutrino as for instance in~\cite{0710.3525,0904.4509,0910.1129,1005.5392,Anderson:2011ns,0811.3424,0812.3661}, which we discuss in section~\ref{sec:rparity}. Whether such a breaking occurs would be subject to a detailed running of the soft-masses from the high-scale which is beyond the scope of this article. 
%%%%%%%%%%%%%%%%%%%%%%%%%%%%%%%%%%%%%%%%%
\subsection{Majorana masses via sneutrino vev and R-parity violation}
\label{sec:rparity}
%%%%%%%%%%%%%%%%%%%%%%%%%%%%%%%%%%%%%%%%%
The alternative to non-perturbatively generated Majorana neutrino masses is to give a vev to the right handed sneutrino which was discussed for example by~\cite{0710.3525,0904.4509,0910.1129,1005.5392,0811.3424,0812.3661}. In principle it is possible to generate the vev for the sneutrino radiatively~\cite{0710.3525,0904.4509,0910.1129,1005.5392} using the running of soft supersymmetric parameters and hence avoiding the problem of D-flatness. As discussed in this literature the sneutrino vev generates a mass for the right-handed neutrino of order the breaking scale and breaks $U(1)_{B-L}\times U(1)$ to hypercharge, breaking the additional $U(1)$ at low energies and leaving only the Standard Model gauge groups. As such this option seems very desirable, however it can lead to dangerous R-parity violating operators depending on the breaking scale and the right-handed neutrino masses. As detailed above the latter are essentially given by the up-quark Yukawa couplings, fixing the breaking scale to the range $10^{11}-10^{13}\, {\rm GeV.}$ Let us illustrate how we can achieve the absence of R-parity violating operators. As discussed in the previous section, we are interested in a solution where $H^{d}_{1}$ has a large supersymmetric mass. Looking at $\langle\tilde{\nu}_{2} \rangle\neq 0$ we then have the following potentially dangerous coupling
\begin{equation}
\mu_{3}H^{u}_{3}.H^{d}_{3}-\frac{v_{2}}{\Lambda}L_{1}H^{u}_{3}\langle \nu_{2}\rangle\, ,
\end{equation}
where the first term is a standard $\mu$ term for the $H_{3}$ Higgs generation and the last term is a bi-linear R-parity violating term $\propto H_{u}L.$ This term can be rotated away by the following field re-definition
\begin{equation}
H^{d'}_{3}=H^{d}_{3}-\frac{v_{2}\langle \nu_{2}\rangle}{\mu_{3}\Lambda}L_{1}\; \text{ and } L^{'}_{1}=L_{1}+\frac{v_{2}\langle \nu_{2}\rangle}{\mu_{3}\Lambda}H^{d}_{3}\, .
\end{equation}
In terms of the rotated superfields and dropping all primes we obtain
\begin{equation}
W=W_{\rm MSSM}-\frac{v_{2}\langle \nu_{2}\rangle}{\mu_{z}\Lambda} \frac{v_{2}\tilde{\phi}\psi}{\Lambda^{3}} L_{2}L_{1}e_{1}-\frac{v_{2}\langle \nu_{2}\rangle}{\mu_{z}\Lambda} \frac{v_{1}\tilde{\phi}\psi}{\Lambda^{3}} Q^{L}_{2}L_{1}d_{1}\, .
\end{equation}
Thus, we have generated lepton number violating terms but not any baryon number generating ones. Constraints from baryogenesis~\cite{0702184} restrict both couplings to be smaller than $10^{-7}.$ The absolute size of these vevs will not be restricted by requiring realistic hierarchical flavour mixings, we still can tune $v_{1,2}/\Lambda,$ only their ratio will be restricted (cf. Equation~\ref{eq:ratios}). With this freedom at hand, we can clearly satisfy the bound R-parity violating operators. Note also that we are attempting to restrict the smallest non-vanishing Yukawa coupling which is already heavily suppressed compared to the other Yukawa couplings. 

%%%%%%%%%%%%%%%%%%%%%%%%%%%%%%%%%%%%%%%%%%
\subsection{Masses and flavour mixing}
\label{sec:mixing}
%%%%%%%%%%%%%%%%%%%%%%%%%%%%%%%%%%%%%%%%%%
The superpotential after breaking becomes:
\begin{eqnarray}
\nonumber W&=&\left(
\begin{array}{c}
 Q_{1}^L \\
 Q_{2}^L \\
 Q_{3}^L
\end{array}
\right)\left(
\begin{array}{ccc}
 0 & H_{3}^u \frac{v_1}{\Lambda} & -H_{2}^u \\
 -H_{3}^u \frac{v_1 \phi \psi}{\Lambda^3} & 0 & H_{1}^u \frac{\phi}{\Lambda} \\
 H_{2}^u \frac{\psi}{\Lambda} & -H_{1}^u & 0
\end{array}
\right)\left(
\begin{array}{c}
 u_{1}^R \\
 u_{2}^R \\
 u_{3}^R
\end{array}
\right)\\
\nonumber &+&\left(
\begin{array}{c}
 Q_{1}^L \\
 Q_{2}^L \\
 Q_{3}^L
\end{array}
\right)\left(
\begin{array}{ccc}
 0 & H_{3}^d \frac{v_1}{\Lambda} & -H_{2}^d \\
 -H_{3}^d \frac{v_1 \tilde{\phi} \tilde{\psi}}{\Lambda^3} & 0 & H_{1}^d \frac{\tilde{\phi}}{\Lambda} \\
 H_{2}^d \frac{\tilde{\psi}}{\Lambda} & -H_{1}^d & 0
\end{array}
\right)\left(
\begin{array}{c}
 d_{1}^R \\
 d_{2}^R \\
 d_{3}^R
\end{array}
\right)\\
\nonumber &+&\left(
\begin{array}{c}
 L_{1} \\
 L_{2}\\
 L_{3}
\end{array}
\right)\left(
\begin{array}{ccc}
 0 & H_{3}^u \frac{v_2}{\Lambda} & -H_{2}^u \\
 -H_{3}^u \frac{v_2 \phi \psi}{\Lambda^3} & 0 & H_{1}^u \frac{\phi}{\Lambda} \\
 H_{2}^u \frac{\psi}{\Lambda} & -H_{1}^u & 0
\end{array}
\right)\left(
\begin{array}{c}
 \nu_{1} \\
 \nu_{2} \\
 \nu_{3}
\end{array}
\right)\\
&+&\left(
\begin{array}{c}
 L_{1} \\
 L_{2} \\
 L_{3}
\end{array}
\right)\left(
\begin{array}{ccc}
 0 & H_{3}^d \frac{v_2}{\Lambda} & -H_{2}^d \\
 -H_{3}^d \frac{v_2 \tilde{\phi} \tilde{\psi}}{\Lambda^3} & 0 & H_{1}^d \frac{\tilde{\phi}}{\Lambda} \\
 H_{2}^d \frac{\tilde{\psi}}{\Lambda} & -H_{1}^d & 0
\end{array}
\right)\left(
\begin{array}{c}
 e_{1} \\
 e_{2} \\
 e_{3}
\end{array}
\right)+W_{\rm D3D7}\, .
\end{eqnarray}
The $D3-D7$ states associated to the breaking fields $\rho_{53}, \Psi_{42}, \Phi_{61}$ get vevs to satisfy D-term equations. As previously discussed we can generate Majorana masses for right handed neutrinos of the following type
\begin{equation}
W_{\rm nm}= A_1 \nu^1_{1}\bar{\nu}_{1} +A_2 \nu_{2} \bar{\nu}_{2} +A_3 \nu_{3}\bar{\nu}_{3}\, , 
\end{equation}
where the $A_i$ are some coefficients which absorb all the sneutrino vevs and other couplings.
For now appropriate configurations for the $\mu-$term are assumed to be present but are not detailed here to focus the analysis on the flavour physics in the quark and lepton sector whose structure we now would like to analyse.
%%%%%%%%%%%%%%%%%%%%%%%%%%%%%%%%%%%%
\subsubsection*{Quark sector}
%%%%%%%%%%%%%%%%%%%%%%%%%%%%%%%%%%%%
The phenomenology of the quark sector is essentially described by the left-right model in \cite{1002.1790}. To get a difference between the down and up-type Yukawa couplings we give the $H_{1}^d$ field a large supersymmetric mass via appropriate $D7$ vevs. To get the correct hierarchical suppression in the CKM matrix we assume the following scaling between the Higgs vevs
\begin{equation}
\frac{H^u_{1}}{H^u_{2}} \sim \epsilon, \qquad \frac{H^u_{3}v_1}{H^u_{2}\Lambda} \sim \epsilon \quad,\quad \frac{\Phi_{61}^u}{\Lambda} \sim \epsilon^{2}\; \mbox{and}  \qquad  \frac{\Phi_{61}^d v_1 H_{3}^d}{\Lambda^2 H_{2}^d} \sim \epsilon. 
\end{equation}
All contributions, including $\Psi_{42}/\Lambda,$ are sub-leading. Then the CKM matrix is approximately
\begin{equation}
|V_{\rm CKM}|=\left(\begin{array}{c c c}
1 & \epsilon & \epsilon^3\\
\epsilon & 1 & \epsilon^2\\
\epsilon^3 & \epsilon^2 &1
\end{array}\right).
\end{equation}
The expression for the mass eigenvalues was presented in \cite{1002.1790} and in the above approximation they lead to the following hierarchies of masses:
\begin{eqnarray}
(m_u^i)^2&=&\left(0,|H_{1}^u|^2+|H_{2}^u|^2\frac{|\Psi_{42}|^2}{\Lambda^2},|H_{2}^u|^2+|H_{3}^u|^2\frac{|\rho_{53}^{(1)}|^2}{\Lambda^2}\right),\\
(m_d^i)^2&=&\left(0,|H_{2}^d|^2\frac{|\Psi_{42}|^2}{\Lambda^2},|H_{2}^d|^2+|H_{3}^d|^2\frac{|\rho_{53}^{(1)}|^2}{\Lambda^2}\right).
\label{eq:quarkmasses}
\end{eqnarray}
The zero mass for the lightest generation is characteristic of models on toric singularities~\cite{1002.1790}. It can be argued to arise from a global symmetry of the low-energy theory which is broken by radiative corrections using bulk effects, leading potentially to realistic masses for the lightest generation~\cite{1102.1973}. 
%%%%%%%%%%%%%%%%%%%%%%%%%%%%%%%%%%%%
\subsubsection*{Lepton sector}
\label{sec:leptonsec}
%%%%%%%%%%%%%%%%%%%%%%%%%%%%%%%%%%%%
Without a mass term for the right-handed neutrinos we would observe the same Yukawa matrices for quarks and leptons at leading order. The difference in the Yukawa matrices coming from the breaking of the $U(4)$ factors is only important at sub-leading order. 
Regarding the leptons this feature is interesting since the individual down-type quarks and lepton masses are not hierarchically different, but it is phenomenologically unrealistic regarding the neutrino and up-type quark masses. The difference might be due to radiative corrections. To leading order the lepton masses are given by
\begin{equation}
m_{e_L}^2= \left(0,|H_{2}^d|^2\frac{|\Psi_{42}|^2}{\Lambda^2},|H_{2}^d|^2+|H_{3}^d|^2\frac{|\rho_{53}^2|^2}{\Lambda^2} \right).
\label{eq:leptonmasses}
\end{equation}

We now turn to the discussion of the seesaw neutrino mass scenario from sneutrino vevs.
Having Majorana masses for right handed neutrino masses we start with the following general mass matrix
\begin{equation}
M_{n}=\left(\begin{array}{c c}
0 & Y\\
Y^T & M_{\rm np}
\end{array}\right).
\end{equation}
This matrix can be brought into block-diagonal form and we then can diagonalise the remaining blocks to obtain the eigenvalues of $\tilde{M}_n.\tilde{M}_n^\dagger$ analytically. We find to leading order the following masses for left and right handed neutrinos:
\begin{eqnarray}
\nonumber m_{\nu_L}^2&=&\left(0, \left(\frac{2 |H_{1}^u|^2}{|A_2|}+\frac{2 |H_{2}^u|^2|\Psi_{42}|^2}{|A_1|\Lambda^2}\right)^2, \left(\frac{2 |H_{2}^u}{|A_3| }+\frac{2 |H_{3}^u|^2 |\rho_{53}^2|^2}{|A_2| \Lambda^2}\right)^2+ \frac{8 |H_{1}^u|^2|H_{3}^u|^2|\rho_{53}^2|^2}{|A_2|\Lambda^2}\right),\\
m_{\nu_R}^2&=& \left(4 |A_1|^2 ,4 |A_2|^2  ,4 |A_3|^2  \right).
\end{eqnarray}
We have the seesaw suppression of left-handed neutrino masses as phenomenologically desired. We note that the left-handed neutrino masses are approximately given by $m_{\nu_L} \sim m_{u,c,t}^2/ m_{\nu_R}.$ Cosmological constraints  imply that $m_{tot} = \sum_i m_{\nu_i} \leq 0.5$~eV. This implies that the right-handed neutrino masses cannot be larger than $\mathcal{O}(10^{13})$~GeV. Further constraints from the SuperK experiment imply that at least one of the left-handed neutrinos should have a mass of at $0.04$~eV. The window for right-handed neutrino masses is then approximately $10^{11} \leq m_{\nu_R} \leq 10^{13}$~GeV.

Next in line is to estimate the flavour mixing in the lepton sector. The flavour mixing is to date known to be present but not as constrained as in the quark sector. The so-called PMMS matrix parametrised by three mixing angles can be written as
\begin{equation}
V_{\rm PMNS}=\left(
\begin{array}{ccc}
 c_{12} c_{13} & c_{13} s_{12} & e^{-i \delta } s_{13} \\
 -c_{23} s_{12}-c_{12} e^{i \delta } s_{13} s_{23} & c_{12} c_{23}-e^{i \delta } s_{12} s_{13} s_{23} & c_{13} s_{23} \\
 -c_{12} c_{23} e^{i \delta } s_{13}+s_{12} s_{23} & -c_{23} e^{i \delta } s_{12} s_{13}-c_{12} s_{23} & c_{13} c_{23}
\end{array}
\right).
\end{equation}
Experimentally the mixing angles are constrained as follows \cite{Amsler:2008zzb}
\begin{equation}
\sin^2{(2\theta_{12})}=0.87\pm 0.03, \; \sin^2{(2\theta_{23})}>0.92, \; \sin^2{(2\theta_{13})}<0.19 \text{ CL}=90\%\, .
\end{equation}
In particular with the undetermined phase $\delta,$ most entries are not heavily constrained from a string model building perspective apart from the fact that there is a large mixing between the first and second generation, which we would like to achieve. As in the quark sector the mixing between the first and second generation is determined (to leading order) by the ratio
\begin{equation}
\frac{H_{3}^d\rho_{53}\Phi_{61}^d}{\Lambda^2 H_{2}^d}\, .
\end{equation}
So far the different vevs of $\rho_{53}$ for lepton and quark sector have not played a role at leading order (e.g.~masses). Here however, we see that the vevs have to differ in order to predict different mixing angles for quarks and leptons. In particular we demand that the ratio of both vevs is similar to
\begin{equation}
\frac{v_1}{v_2}\sim \epsilon^2\, .
\label{eq:ratios}
\end{equation}

 The size of this ratio $v_1/v_2$ also implies that the mixing angle between the first and third generation for leptons is smaller than the corresponding angle for the quarks. This can be seen from the scaling of the (1,3) entry in the PMNS and which is given by
\begin{equation}
\sin{\theta_{13}}^{PMNS}=\frac{\epsilon^3}{\sqrt{1+\left(\frac{H_{3}^d\rho^{(2)}_{53}\Phi_{61}^d}{\Lambda^2 H_{2}^d}\right)^2}}\, ,
\end{equation}
which includes a factor of $v_2=\rho_{53}^{(2)}$ in the denominator. The corresponding angle in the CKM matrix instead has $v_1=\rho_{53}^{(1)}$ and since and since $v_1<v_2$ the mixing angle is smaller for leptons. Our model thus does not display tribimaximal mixing. Recent results from the T2K experiment~\cite{Abe:2011sj} have indicated a non-zero value for this angle, $0.08< \sin{\theta_{13}}^{PMNS} < 0.27$ being the 90\% confidence limits. Since $\sin{\theta_{13}}^{CKM}\approx 0.0035$ there is a slight tension between our prediction and the T2K result. However, we expect that due to radiative corrections necessary to generate the first family fermion masses that this may also change. Also the remaining mixing angle depends on the size of the right-handed neutrino masses which is not controlled within the local construction.

%%%%%%%%%%%%%%%%%%%%%%%%%%%%%%%%%%%%%%%%%
\subsection{Counting parameters}
%%%%%%%%%%%%%%%%%%%%%%%%%%%%%%%%%%%%%%%%%

Neglecting the complex phases for now, where we expect some relative alignment arising from the scalar potential, we have 9 parameters (6 masses + 3 angles) in the quark sectors and 12 parameters (9 masses + 3 angles) in the lepton sector, totalling 21 parameters. In our model we choose 15 non-vanishing vevs as summarised in Table~\ref{listofvevs}.
\begin{center}
\begin{tabular}{c |c| c| c| c| c| c}
$\langle\rho_{35}\rangle$ & $\langle\Phi_{61}\rangle$ & $\langle \Psi_{42} \rangle$ & $\langle H^{u,d}_{1}\rangle$ & $\langle H^{u,d}_{2}\rangle$ & $\langle H^{u,d}_{3}\rangle$ & $A_i$  \\ \hline
2 & 2 & 1 & 2 & 2 & 2 & 3 
\end{tabular}
\captionof{table}{\footnotesize{A list of parameters that we have to adjust to be non-vanishing in our model to achieve for Standard Model masses and mixing angles.}\label{listofvevs}}
\end{center} 
Our model gives expressions for the 21 SM parameters in terms of these 15 parameters. The fact that there are more SM parameters than model parameters implies that there must be relations between the expressions for the SM parameters. Some of these are clear; for instance $m_u=m_d=m_e=m_{\nu_1}=0$. There is also a relation between the quark and neutrino masses, $m_{\nu_L} \sim m_{u,c,t}^2/m_{\nu_R}$.  Finally there are relations between some of the light quarks and leptons, which we do not consider robust predictions of our model as they will be changed during compactification~\cite{1102.1973}.

%%%%%%%%%%%%%%%%%%%%%%%%%%%%%%%%%%%%%%%%%
\subsection{Proton decay}
\label{sec:protondecay}
%%%%%%%%%%%%%%%%%%%%%%%%%%%%%%%%%%%%%%%%%
Here we discuss proton decay operators up to dimension six and find that the additional (anomalous) $U(1)$ symmetries severely restrict the possible operators. Our analysis is based on the list of proton decay operators in Table~3 of~\cite{9908305}.

Before forbidding potential operators with the anomalous $U(1)$ symmetries, let us recall how we can break these $U(1)$ symmetries for which we have two possibilities already used in this article:
\begin{enumerate}
\item Using non-perturbative corrections arising for example from $E3$ branes wrapping a 4-cycle intersecting with the singularity (cf. the $\mu-$term ). In this case a 4-cycle K\"ahler modulus is charged under the $U(1)$ symmetry and can generate, given sufficient zero mode structure, the perturbatively forbidden coupling in the superpotential. 
\item A vev for the scalar component of a bi-fundamental field charged under two anomalous $U(1)$ symmetries (cf. breaking to the Standard Model gauge groups and a vev for the right-handed sneutrino). Couplings breaking these $U(1)$ symmetries are couplings which involve the field breaking the symmetry. No further couplings that can either be constructed without the breaking field, or be constructed by integrating out the breaking field are induced. Hence for example the dangerous coupling $udd\nu_{R}/\Lambda$ is forbidden by the $U(1)$ symmetries that are broken by the vev of $\nu_{R}$ and hence also cannot be induced if the right-handed sneutrino obtains a vev. On the contrary a higher dimensional Yukawa coupling of the type $\frac{\Phi_{61}}{\Lambda} Q_{L} H_{u} u_{R}$ remains present in the low-energy theory.
\end{enumerate}

The $U(1)_{R}$ symmetry in the superpotential Eq.~\ref{tab:globalsymmetries} restricts the allowed couplings such that higher powers of a given superpotential term are forbidden at tree level. The non-renormalisation of the superpotential at the perturbative level then guarantees that these operators are not induced perturbatively. Hence dangerous operators that one might imagine involving $D3-D7$ or $D7-D7$ states are simply absent in the superpotential at the perturbative level. On the non-perturbative level and in the K\"ahler potential, as highlighted in Table~\ref{tab:protdecay}, the remaining operators are mostly forbidden by $U(1)_{B-L}.$ The only surviving operators are $Q_{L}Q_{L}Q_{L}Q_{L}$ and $Q_{R}Q_{R}Q_{R}Q_{R}$ as potential non-perturbatively generated operators in the superpotential. All other operators in the K\"ahler potential are forbidden. However the operators $Q_{L}Q_{L}Q_{L}Q_{L}$ and $Q_{R}Q_{R}Q_{R}Q_{R}$ are not generated with the Euclidean 3-brane setup we choose in our model. In the absence of a global completion of our local model, one cannot say whether such operators are induced by other more complicated non-perturbative effects. Such effects are beyond the scope of local model building. In summary, the R-symmetry and the $U(1)_{B-L}$ forbid proton decay arising from both D- and F-term operators sufficiently.

\begin{table}\begin{center}
\begin{tabular}{|c|c|c|}
\hline
      Operator &  Dimension & Forbidden by
 \\ \hline\hline
$[QQQL]_F$ & 5 & anomalous $U(1)$   \\ \hline
$[uude]_F$ & 5 & anomalous $U(1)$    \\ \hline
$[QQQH]_F$ & 5 & $U(1)_{B-L}$ \\  \hline\hline
$[\bar H\bar H e^*]_D$ & 5 & $U(1)_{B-L}$ \\  \hline
$[QuL^*]_D$ & 5 & $U(1)_{B-L}$  \\ \hline
$[\bar H^* H e]_D$ & 5 & $U(1)_{B-L}$ \\ \hline
$[QQd^*]_D$ & 5 & $U(1)_{B-L}$  \\ \hline  \hline
$[uuuee]_F$ & 6 & $U(1)_{B-L}$  \\ \hline
$[uddH\bar H]_F$ & 6 & $U(1)_{B-L}$  \\ \hline
$[dddLH]_F$ & 6 & $U(1)_{B-L}$  \\ \hline
$[uddL\bar H]_F$ & 6  & $U(1)_{B-L}$ \\ \hline \hline
$[AA^*LH^*]_D$ & 6  & $U(1)_{B-L}$ \\ \hline
$[AA^*L\bar H]_D$ & 6  & $U(1)_{B-L}$ \\ \hline
$[QQQ\bar H^*]_D$ & 6  & $U(1)_{B-L}$ \\ \hline
$[QQu^*e^*]_D$ & 6 & $U(1)_{B-L}$  \\ \hline
$[Qu^*d^*H]_D$ & 6 & $U(1)_{B-L}$   \\ \hline
$[Qu^*d^*L]_D$ & 6 & $U(1)_{B-L}$  \\ \hline
$[Qu^*d^*\bar H^*]_D$ & 6 & $U(1)_{B-L}$  \\ \hline
$[Qd^*d^*\bar H]_D$ & 6 & $U(1)_{B-L}$  \\ \hline
$[Qd^*d^*L^*]_D$ & 6 & $U(1)_{B-L}$  \\ \hline
$[Qd^*d^*H^*]_D$ & 6 & $U(1)_{B-L}$ \\ \hline
$[Qu\bar H^* e]_D$ & 6 & $U(1)_{B-L}$ \\ \hline
$[QdH^*e^*]_D$ & 6 & $U(1)_{B-L}$  \\ \hline
$[Qd\bar H e^*]_D$ & 6 & $U(1)_{B-L}$  \\ \hline
$[LLH^*H^*]_D$ & 6 & $U(1)_{B-L}$  \\ \hline
$[ddde^*]_D$ & 6 & $U(1)_{B-L}$ \\   \hline
\end{tabular}
\captionof{table}{\footnotesize{Summary of proton decay operators of dimension 5 and 6 taken from Table 3 of~\cite{9908305}. $A$ denotes any possible MSSM field. $e$ represents either the right-handed electron or neutrino in our model. Besides being forbidden by $U(1)_{B-L},$ which for us seems the strongest constraint, anomalous $U(1)$ symmetries restrict all couplings.}\label{tab:protdecay}}
\end{center}
\end{table}

%%%%%%%%%%%%%%%%%%%%%%%%%%%%%%%%%%%%%%%%%%%%%%%%%%
\subsection{Gauge coupling unification}
\label{sec:gaugecoupling}
%%%%%%%%%%%%%%%%%%%%%%%%%%%%%%%%%%%%%%%%%%%%%%%%%%

We now turn to investigating gauge coupling unification in our model. At the high scale the gauge coupling for the non-abelian gauge groups is given by the tree-level value of the dilaton. 
With gauge coupling unification implemented at the high scale the question to ask is whether we can reproduce the observed values for the inverse gauge coupling at the weak scale $(\alpha_1^{-1}=98.33\pm 0.03,$ $\alpha_2^{-1}=29.57\pm 0.03,$ $\alpha_3^{-1}=8.45\pm 0.05)$ at $M_Z=91.1876\pm 0.0021\,{\rm GeV}$ \cite{Amsler:2008zzb,0908.1135}. In our analysis we restrict ourselves to one loop beta functions, neglect threshold effects and restrict ourselves to completely supersymmetric running for simplicity. 

The one-loop beta function for a supersymmetric theory with a product gauge group $G_1\times G_2$ is given by~\cite{Jones:1981we}
\begin{equation}
\beta_{g_1} = T(R_1) d(R_2) - 3C_2(G_1)\, ,
\label{eq:jones}
\end{equation} 
where for an irreducible matrix representation $R^a$, $T(R)$ is defined by $Tr[R^a R^b] = T(R) \delta^{ab}$ (and from now on $T(R)=1/2),$ $d(R)$ is the dimension of the representation (the dimension of the fundamental of $SU(N)$ is $d(\square_N)=N)$ and $C_2(G)$ is the quadratic Casimir of the adjoint representation (note that $C_2(SU(N))=N$). A sum over all field content in the above formula is implicit.  We use Eq.~\ref{eq:jones} to compute the beta functions for any gauge groups that can appear in the breakdown from the $dP_3$ gauge groups at the string scale down to the MSSM gauge groups at low energies. Here we keep the most general breaking pattern possible
\begin{eqnarray}
SU(4)_a\times SU(4)_b &\overset{M_{4d}}{\to}&SU(4)\overset{M_{4}}{\to}SU(3)\times U(1)_{B-L}\, ,\\
 SU(2)_{La}\times SU(2)_{Lb}&\overset{M_{2Ld}}{\to}&SU(2)_L\, ,\\
 SU(2)_{Ra}\times SU(2)_{Rb}&\overset{M_{2Rd}}{\to}&SU(2)_R \overset{M_Y}{\to} U(1)\, ,
\end{eqnarray}
where the mass scale $M_X$ denotes the breaking scale of that gauge group. In addition to the above breaking scales we allow for two further variable scales, the scale the additional Higgs fields become massive $M_{\rm higgs}$ and the scale the $D7$ gauge content becomes massive $M_{D7}.$

Recall that the running of the inverse gauge coupling is given by
\begin{equation}
\alpha_{X}^{-1}(M)=\alpha_X^{-1}(M_X)+\frac{\beta_X}{2\pi}\log{\left(\frac{M_X}{M}\right)}\, .
\end{equation}
At the breaking scale $M_X$ there are matching conditions for the gauge couplings which depend on the normalisation of the  unbroken group with respect to the  larger gauge group before the breaking. A detailed analysis of these matching conditions and the running of the inverse gauge couplings is given in Appendix~\ref{sec:running}. We summarise in Figure~\ref{fig:couplingbreakdown} the gauge couplings valid in between individual breaking scales. 
\begin{center}
\begin{tabular}{c c}
\begin{tabular}{c}\includegraphics[width=0.55\textwidth]{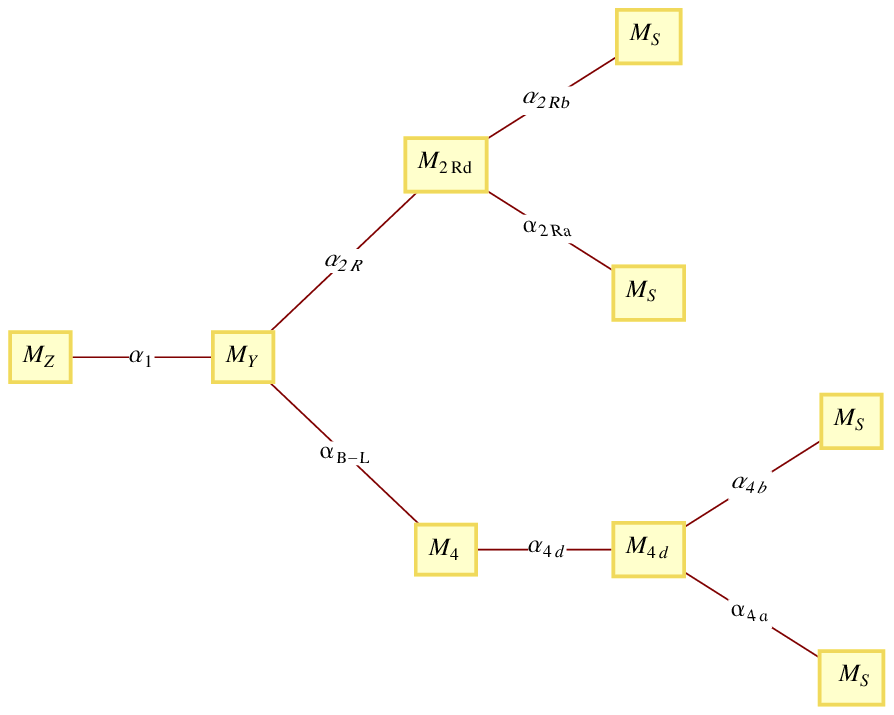} \end{tabular}& \begin{tabular}{c} \includegraphics[width=0.35\textwidth]{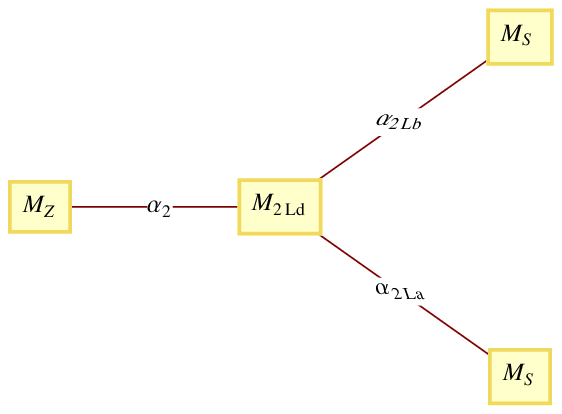}\\\includegraphics[width=0.45\textwidth]{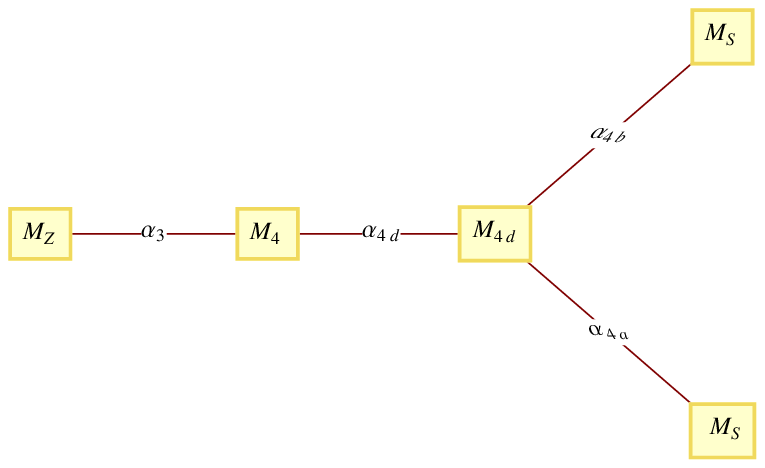}\end{tabular}
\end{tabular}
\captionof{figure}{{\footnotesize{On the left side the breakdown from the $dP_3$ gauge groups to $U(1)_Y$ is shown. The vertices denote the associated breaking scales. At each of these breaking scales we need to apply the matching condition from Equation~\ref{boundarydiagonal} for the inverse gauge couplings.}}\label{fig:couplingbreakdown}}
\end{center}
We find the following bound on the string scale, looking at the difference between the $SU(3)$ gauge coupling and the $U(1)_Y$ gauge coupling at $M_Z$:
\begin{eqnarray}
\nonumber \alpha^{-1}_Y(M_Z)-\alpha^{-1}_3(M_Z)=89.88&\leq& \frac{4}{3}\alpha^{-1}(M_s)+\frac{8}{\pi}\log{\left(\frac{M_s}{M_Z}\right)}\\
\Rightarrow M_s &\gtrsim & 10^{14}\, {\rm GeV}\, ,\label{lowerbound}
\end{eqnarray}
where in the last estimate we neglect the influence of the term involving $\alpha^{-1}(M_s).$ The details of this estimate can be found in Appendix~\ref{sec:couplingestimate}. We remind the reader that the unification scale $M_{s}$ does not have to be the string scale but can be the string scale enhanced by the radius of the overall compactification $M_{UV}\sim M_{s}{\cal V}^{1/6}$ \cite{0901.4350,0906.1920}. The above upper bound refers to the unification scale $M_{UV}$ and hence the upper bound on the string scale is lower. We can achieve for a string scale at the intermediate scale which depending on the structure of soft masses \cite{1003.0388,1011.0999} can be consistent with TeV soft-masses.

Using the running derived in Appendix~\ref{sec:running}, we find the following notable scenarios:
\begin{enumerate}
\item {\bf Intermediate string scale:} Breaking the copies of $SU(4)$ and $SU(2)_R$ at the string scale, we find a scenario with gauge coupling unification, which saturates the lower bound on the string scale (cf. Equation~\ref{lowerbound}) of $10^{14}$~GeV. Note that for now the additional Higgs generations are allowed to survive until the weak scale. Bounds on flavour changing neutral currents generally forbid these but only to the PeV scale~\cite{1002.0900}, changing the nature of the running not significantly in this approximation. The running is illustrated in Figure~\ref{runninglowstringscale}.
\begin{center}
\includegraphics[width=0.7\textwidth]{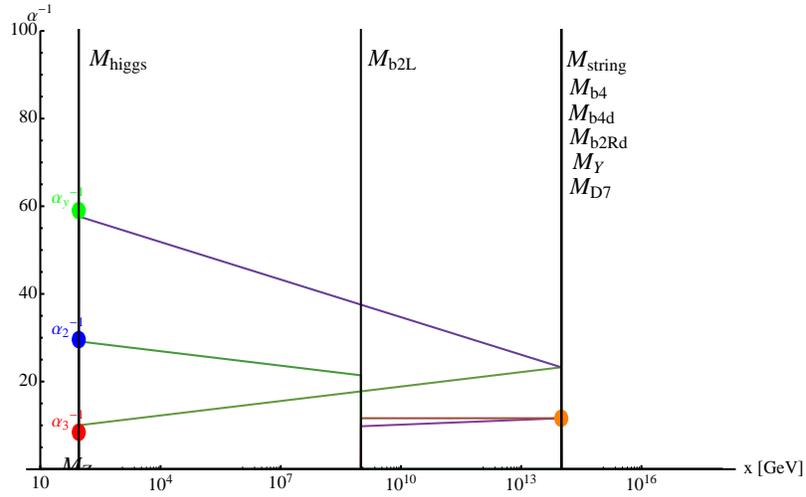}
\captionof{figure}{\footnotesize{Shown is the running for the inverse gauge couplings applicable in between the relevant scales. The green, blue and red circles at $M_Z$ denote the experimentally observed values for the gauge coupling constants. The orange circle at $M_s$ denotes the inverse gauge coupling at the string scale.}\label{runninglowstringscale}}
\end{center} 
\item {\bf GUT scale string scale:} This is an example of a string scale that is close to the GUT scale at $M_s=10^{16}\,{\rm GeV}.$ The breakdown to the MSSM gauge group is at the high scale. The running is illustrated in Figure~\ref{runninghighstringscale}.
\begin{center}
\begin{tabular}{c c}
\includegraphics[width=0.45\textwidth]{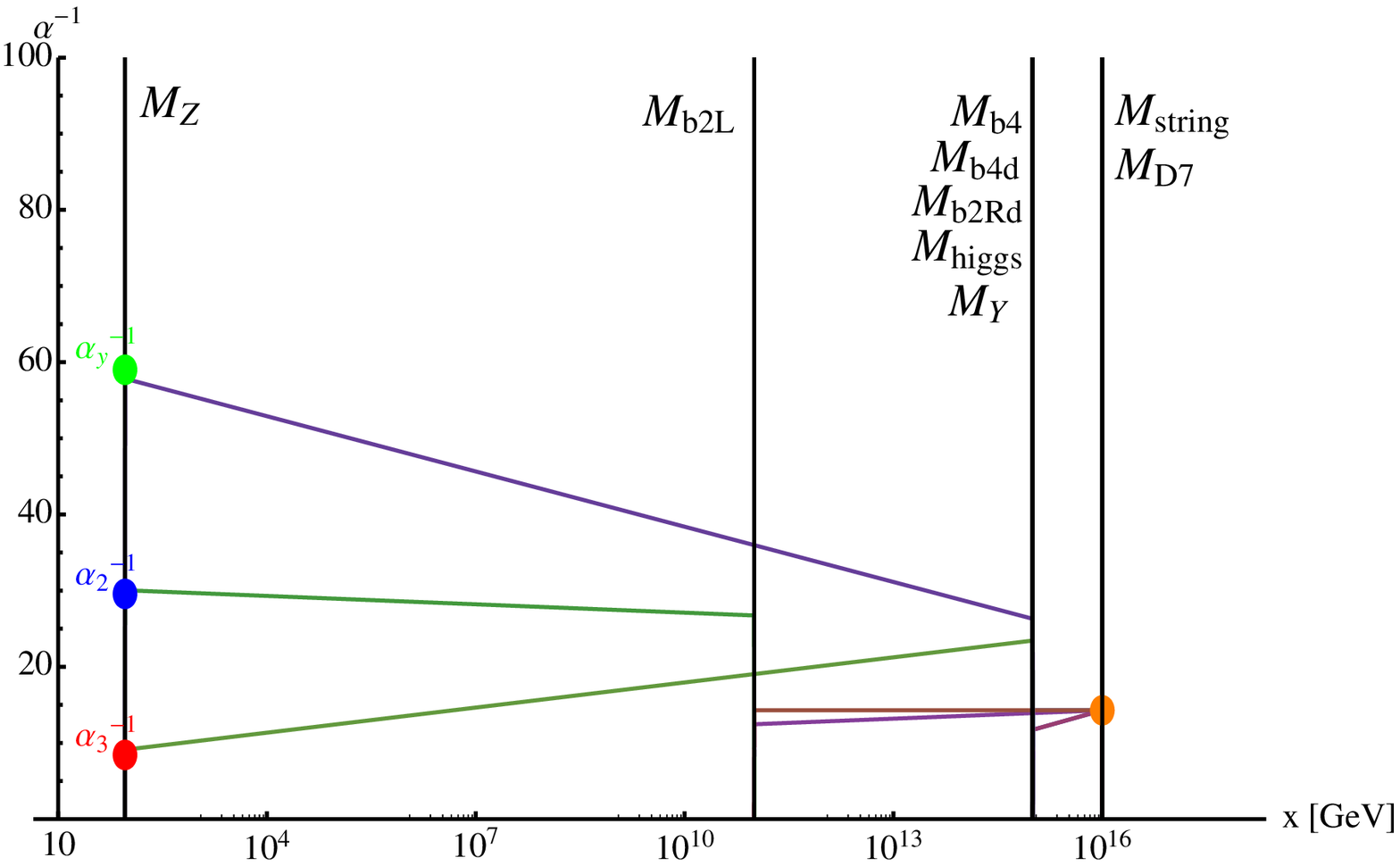}&\includegraphics[width=0.45\textwidth]{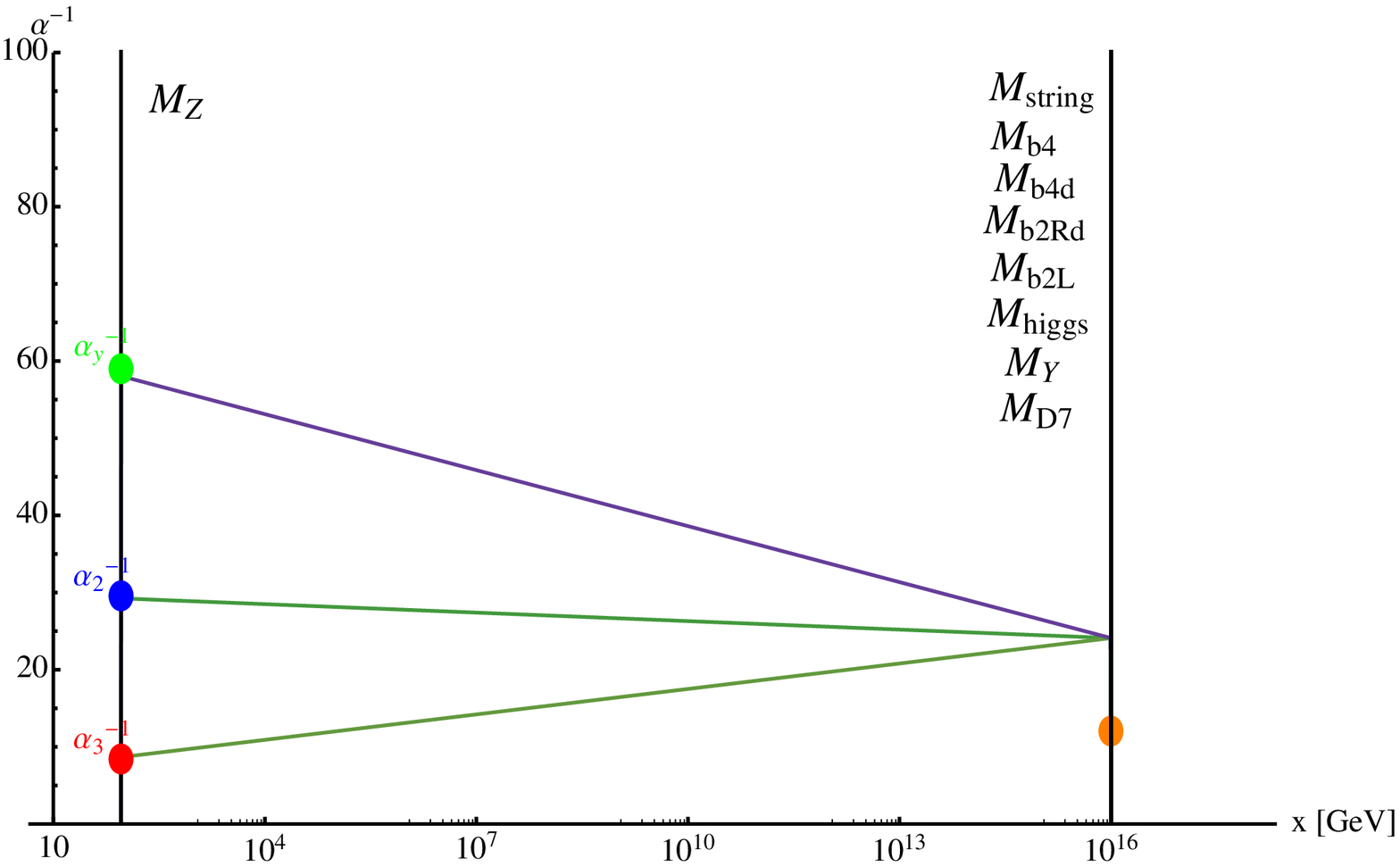}
\end{tabular}
\captionof{figure}{\footnotesize{In similar fashion to the intermediate string scale scenario we show the running for the inverse gauge couplings. In the scenario on the left, the breaking down to the Standard Model gauge groups takes place at two stages, whereas the breaking in the scenario on the right takes place at the string scale at $10^{16}$ and we observe the standard gauge coupling unification of the MSSM.}\label{runninghighstringscale}}
\end{center}
\item {\bf Additional left-handed $W$ bosons at the LHC:} The breaking of the two $SU(2)_{L}$ gauge groups can be as low as the TeV scale from the perspective of gauge coupling unification, which offer interesting phenomenological opportunities in the LHC era. The running is visualised below in Figure~\ref{runninglowws}.
\begin{center}
\includegraphics[width=0.7\textwidth]{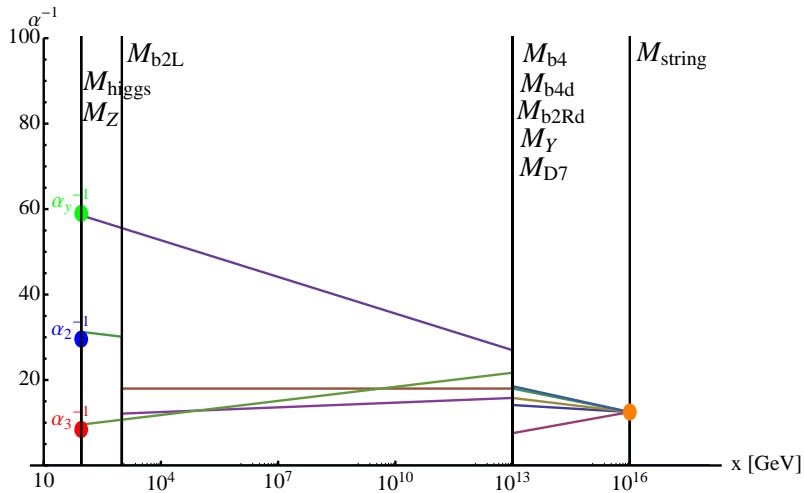}
\captionof{figure}{\footnotesize{Shown is the running for the inverse gauge couplings applicable in between the relevant scales. The green, blue and red circles at $M_Z$ denote the experimentally observed values for the gauge coupling constants. The orange circle at $M_s$ denotes the inverse gauge coupling at the string scale.}\label{runninglowws}}
\end{center} 
\end{enumerate}

%%%%%%%%%%%%%%%%%%%%%%%%%%%%%%%%%%%%%%%%%%%%
\section{Consistency of scales}
%%%%%%%%%%%%%%%%%%%%%%%%%%%%%%%%%%%%%%%%%%%%
After discussing the model, we now comment on the consistency of scales associated to the breakdown to the Standard Model gauge symmetries and gauge coupling unification with the scales of  moduli stabilisation and supersymmetry breaking. In particular we would like to compare it with the two supersymmetry breaking scenarios in the context of Large volume compactifications~\cite{0502058}, depending on whether moduli redefinitions occur~\cite{1003.0388,1011.0999} or not~\cite{0906.3297}. Hierarchies are created by different suppressions in terms of the overall bulk volume in string units ${\cal V}.$ In both scenarios we control contributions to the scalar potential up to $1/{\cal V}^3.$ Hence we demand D- and F-flatness up to that order. We have to ensure that any contribution from the matter fields is sub-leading, and discuss in turn the requirements. The different scenarios for soft-masses essentially fix the overall bulk volume to provide a 'solution' to the hierarchy problem. In scenario 1 with soft masses of order $m_{\rm soft}\sim 1/4\pi^2 M_{\rm P}/{{\cal V}}$ we find a volume ${\cal V}\sim 10^{12}.$ In scenario 2 the smallest soft masses scale as $m_{\rm soft}\sim  M_{\rm P}/{{\cal V}^2}$ which then corresponds to a volume of order ${\cal V}\sim 10^{7}.$ 

These soft-masses are obtained when starting with the following~\cite{0609180,0805.2943} tree-level scaling in the matter K\"ahler potential
\begin{equation}
K_m=\frac{1}{{\cal V}^{2/3}}\phi_i\phi_i^\dagger\, ,
\end{equation}
where $\phi_i$ denotes any matter field.
We satisfy the constraint from the contribution to the potential by requiring the fields to obtain vevs less than of the order
\begin{equation}
\langle \phi^c \rangle\leq \frac{M_P}{{\cal V}^{3/4}}\,.
\label{minsupp}
\end{equation}
In the largest case, the F-term of the matter field
$|F^\phi|=|K_{\phi\bar{\phi}}F^\phi \bar{F}^{\bar{\phi}}|^{1/2}$ scales as $|F^\phi|\sim M_P/{\cal V}^{3/2}.$

In scenario 1, this limiting case of a suppression as in Equation~\ref{minsupp} is consistent with the smallness of the matter F-terms as required by the other soft-masses. Using a volume of order ${\cal V}\sim 10^{12},$ we hence have a maximal breaking scale of order
\begin{equation}
\langle\phi^c\rangle\sim M_b\sim 10^{9}\,\rm{GeV}.
\end{equation}
In this setup, unification shall occur at $M_{UV}\sim M_P/{\cal V}^{1/3}\sim 10^{14}\,{\rm GeV.}$ At the superficial level with the exclusion of threshold effects and supersymmetry breaking effects in the running, we find the breaking at $10^9~{\rm GeV}$ too small for gauge coupling unification. We note that the inclusion of warping effects or tuning $W_0$ might evade this constraint.

In the second scenario, the F-terms of the matter fields have to be suppressed up to $1/{\cal V}^2$ and hence require the matter fields to obtain a vev of the order
\begin{equation}
\langle\phi^c\rangle\sim M_b\sim \frac{M_P}{{\cal V}}\sim 10^{12}\,\rm{GeV},
\end{equation}
where we assumed a volume of the order ${\cal V}\sim 10^6.$ As shown in Figure~\ref{runninglowws} we find this limiting scale to be consistent with gauge coupling unification.

%%%%%%%%%%%%%%%%%%%%%%%%%%%%%%%%%%%%%
\section{Conclusions and Outlook}
%%%%%%%%%%%%%%%%%%%%%%%%%%%%%%%%%%%%%
We have studied the construction of realistic models on del Pezzo singularities and argued that in order to account for the hierarchies of masses, flavour structure and symmetry breaking patterns $dP_0, dP_1$ and $dP_2$ are not rich enough whereas $dP_3$ satisfies all the requirements.  We presented a brane model based on the third del Pezzo singularity $dP_3$ that gives rise to all Standard Model fermions with three pairs of Higgs-doublets at low-energies with realistic Yukawa structure for both quarks and leptons. The control over the flavour sector can arise through the control of the leading order K\"ahler potential for the matter fields, utilising the additional gauge structure of anomalous $U(1)$ symmetries in $dP_3.$ These $U(1)$ symmetries along with $U(1)_{B-L}$ play a crucial role in forbidding proton decay. To achieve the desirable flavour parameters (12 masses and 6 angles) we tune 14 parameters.  We use a vev for the right-handed sneutrino to generate Majorana neutrino masses and checked that R-parity violation be sufficiently suppressed. Our model also predicts that the angle $\theta_{13}$ of the PMNS mixing matrix is non-zero but detailed values may be affected by loop effects. 

Depending on the scale of breakdown to the Standard Model gauge groups we found gauge coupling unification in the energy range from $M_{\rm UV}\sim 10^{14}\, {\rm GeV}$ (which can correspond to  a string scale at $M_s\sim 10^{12}\,{\rm GeV})$ to the usual unification scale at $M_{\rm UV}\sim 10^{16}\, {\rm GeV},$ without the requirement of one GUT group at high energies. A priori, the lower bound on the string scale sounds very interesting in the context of standard soft-masses~\cite{0906.3297} being corrected by loop-effects to the order of the gravitino mass divided by loop suppression factors~\cite{1003.0388,1011.0999} since this still can lead to TeV scale soft-masses. Unfortunately, the breaking scale of the additional gauge symmetries to the Standard Model is not unrelated to supersymmetry breaking since F-flatness is only guaranteed to the order the potential is controlled. This upper bound on the breaking scales renders it impossible to achieve gauge coupling unification, without the inclusion of threshold effects and supersymmetry breaking effects, at an intermediate string scale unless tuning $W_0$ or introducing warping effects. However, the upper bound on the breaking scale is found to be consistent with a string scale near the usual unification scale at $10^{16}\,{\rm GeV}$ and TeV soft-masses as in the scenario presented in~\cite{0906.3297}. A proper discussion of supersymmetry breaking terms is beyond the scope of this article.

These detailed properties distinguish the model on $dP_3$ from previously presented models as for example in~\cite{0001083} and~\cite{0508089}. In particular we should list the realistic Lepton Yukawa couplings, realising hypercharge from non-abelian factors hence guaranteeing it to be massless, the complete scenarios for flavour physics and unification presented as successes of the construction on $dP_3.$ 

Depending on the breaking scales and the favourite model of unification, we find very interesting phenomenological signatures of our model, naively in the reach of LHC physics, such as additional $U(1)$ symmetries, additional $SU(2)$-bosons and an interesting Higgs sector.

However, at this stage of model building there still remain some open problems offering rich prospects for future work:
\begin{itemize}
\item To achieve the desirable flavour and mass properties, we need to tune several vevs by hand. It remains open to determine the potential for these vevs and to ask whether these vevs actually can be achieved dynamically.
\item To explain via a concrete model, possibly in a concrete compact or semi-local setup, how the $D3-D7$ states get masses via $D7-D7$ states such that they are not present at low energies.
\item To find an explicit realisation of the scenario with intermediate scale string scale and unification which required warping or tuning of the flux parameter $W_0.$
\item To obtain the correct scale for neutrino masses dynamically.
\item To find an embedding of this model on $dP_3$ in an honest-to-God string compactification in terms of a compact Calabi-Yau with $dP_3$ singularities.
\end{itemize}
The list of achievements and open questions illustrates that the bar on realistic string models is increasing with time which is encouraging. We hope to return to some of these questions in future work.

\section*{Acknowledgments}
It is a pleasure to thank Ben Allanach, Cliff Burgess, Joe Conlon, Luis Ib\'a\~nez, Sadia Khalil, Anshuman Maharana, Noppadol Mekareeya, Gary Shiu, Pablo Soler, Angel Uranga, and Timo Weigand for useful discussion. SK would like to thank the Abdus Salam Centre for Theoretical Physics, the HKIAS and in particular Gary Shiu for hospitality where part of this work was completed. MJD wishes to thank CERN for hospitality during the completion of this work.

\appendix
%%%%%%%%%%%%%%%%%%%%%%%%%%%%%%%%%%%%%%%%%%%%%%%%%
\section{Consistency of Breaking}
\label{sec:consistency}
%%%%%%%%%%%%%%%%%%%%%%%%%%%%%%%%%%%%%%%%%%%%%%%%%

%%%%%%%%%%%%%%%%%%%%%%%%%%%%%%%%%%%%%%%%%%%%%%%%%
\subsection{D-Flat}
%%%%%%%%%%%%%%%%%%%%%%%%%%%%%%%%%%%%%%%%%%%%%%%%%
After vevving $D3-D3$ fields such as $\Phi_{61}$ to break down to the Standard Model, we must ensure that the supersymmetric D-terms are satisfied. The abelian D-term equations can always be satisfied by tuning the FI terms and they do not require further consideration.
The non-abelian D-terms must be satisfied at each node of the quiver, and this requires that also $D3-D7$ fields obtain vevs. To demonstrate this, consider the non-abelian D-term conditions at the $SU(4)$ node, letting $t^a$ be the $SU(4)$ generators. If we denote the incoming fields as $X^p$ and the outgoing fields as $Y^q$, the following condition must be satisfied:
\begin{equation}
\sum_p \bar{X}^p t^a X^p = \sum_q \bar{Y}^q t^a Y^q\, .
\label{eq:dterm}
\end{equation}
Similar equations hold at all other nodes of the quiver. If no fields were vevved, Equation~\ref{eq:dterm} would
certainly be satisfied in the vacuum of the theory. It should also be clear that if an incoming field has a vev, then an outgoing field will also need a vev to satisfy the equation. As all fields in the quiver are bi-fundamentals, this vevved outgoing field will be the ingoing field for another node of the quiver whose D-term condition must be satisfied. Continuing to apply this reasoning around the quiver, we see that for the D-terms to be satisfied the vevved fields must form closed cycles in the quiver, and hence correspond to gauge invariant operators \cite{Buccella:1982nx,9506098}.
We can therefore satisfy all the D-term equations by vevving $37$ string states associated with the vevved $D3-D3$ state.

One can easily check that the D-flat conditions for the VEV $\langle \rho_{53}\rangle_{1}$ are satisfied but they are not satisfied for $\langle \rho_{53}\rangle_{2}$ (all are satisfied apart from $T^{15}).$ An additional VEV is needed which for example is given by a pair of $D3-D7$ states, call them $\rho_{7a}$ and $\rho_{7b}.$ The same applies to $\langle\Phi_{61}\rangle.$ In particular the vev in Equation \ref{rho7vev} and \ref{phi7vev} can be used to satisfy the non-abelian D-term equations, leading to conditions that can easily be satisfied
\begin{equation}
0=-|v_{1}|^{2}+|v_{2}|^{2}+|\rho|^{2}, \text{ and }0=|\phi|^{2}-|\tilde{\phi}|^{2}+|\phi_{7}|^{2}\, .
\end{equation}

\subsection{F-flat}
\label{sec:fflat}
To satisfy D-flatness, we had to vev $D3-D7$ states. Now one can check that the cubic coupling among the $D3-D3$ and the $D3-D7$ states is no longer F-flat in the global supersymmetric sense $\partial_{i}W\neq 0.$ However we are working in an effective supergravity framework, implying that we should look at the associated F-term in supergravity meaning $F^{i}=K^{i\bar{j}}D_{\bar{j}}\bar{W}\neq 0,$ where $D_{i}W=\partial_{i}W+W\partial_{i}K$ denotes the usual covariant derivative in supergravity. In the effective supergravity we have only control over the (next-to)-leading order contribution to the potential and in this framework we can trust, respectively have to guarantee, F-flatness up to the order we can trust our effective supergravity theory. For concreteness we work in a large volume effective supergravity~\cite{0502058} where the expansion parameter is the overall bulk volume and beyond the leading order contribution to the scalar potential, corrections in $g_{s}$ and $\alpha'$ ruin the approximation. The effective supergravity setup is given by
\begin{eqnarray}
K&=&-2\log{\left(\cal V+\xi\right)}+\frac{f(\tau_{i})}{{\cal V}^{2/3}}\left(\sum_{j}\Phi_{j}\Phi_{j}^{\dagger}\right)\, ,\\
W&=&W_{0}+A e^{-aT_{s}}W_{\rm matter}\, ,
\end{eqnarray}
where ${\cal V}$ denotes the overall volume of the bulk geometry, $\xi$ the leading order $\alpha'$ corrections and $\tau_{i}$ 4-cycle volumes, $\Phi_{i}$ denotes all matter fields, $W_{0}$ the flux parameter. For further details we refer the reader to the review literature on the subject~\cite{0611039,0907.0665}. In this supergravity setup we can show that depending on the size of the vev for the matter fields, we can achieve F-flatness. Assuming an overall scaling of $\langle \Phi_{i}\rangle\sim M_{P}/{\cal V}\sim m_{3/2},$ we find that all F-terms are suppressed at least at $F^{i}\sim 1/{\cal V}^{7/3}.$ This is an additional suppression compared to the non-vanishing F-terms of the K\"ahler moduli by more than $1/{\cal V}$ which for us is sufficient for F-flatness.  Having a larger suppression for the vev of the field would render the F-term even smaller.

\subsection{Masses}
A vev for any bi-fundamental field can induce masses via the D-term couplings. Taking for example the diagonal vev for $\Psi_{42}$ as in Equation~\ref{eq:breaking2} the induced masses are proportional to $m_{\psi}\sim g \psi.$ Out of the original eight real degrees of freedom in $\Psi_{42},$ all of them apart from four, which are eaten as Goldstone bosons of the broken $U(2),$ obtain a mass. The same can be checked explicitly in the $SU(4)$ case. The $D3-D7$ states do not obtain a mass via the $U(2)$ D-term but obtain a mass via couplings with $D7-D7$ couplings which are not part of the local construction. We conclude that after breaking all degrees of freedom in $\Psi_{42}$ become massive as desired. In a similar way it can be shown that all degrees of freedom involved in the breaking of the $SU(4)$ and $SU(2)_R$ become massive.

\section{Non-anomalous $U(1)$ symmetries}
As for example discussed in \cite{9909172}, the non-anomalous $U(1)$ symmetries can be identified as follows: The anomaly of interest is $U(1)_i\times SU(N_j)^2.$ The anomalies are best summarised in an $n\times n$ matrix whose entry $(i,j)$ corresponds to the anomaly between $U(1)_i\times SU(N_j)^2.$ The entry $T_{ij}$ is given by
\begin{equation}
T_{ij}=\sum_{k=1}^{m}\sum_{l=1}^N \#_k M^{U(1)}_{lj}\delta\left(M^{SU(N)}_{li}\right)M^{SU(N)}_{lk}\, ,
\label{anfree}
\end{equation}
where $M^{U(1)}$ is a matrix that includes the $U(1)$ charges for all fields, $M^{SU(N)}$ is the matrix with all charges for $SU(N)$ gauge groups, $\#$ indicates the multiplicity of each field and $\delta(.)$ returns $1$ only if that matrix element is non-zero. Generally one then finds that all $U(1)$ symmetries will be anomalous for itself but that there are linear combinations of $U(1)$ symmetries which will be anomaly-free. Here we are interested in linear combinations that involve only combinations of $D3$ gauge groups since in general the $D7$ gauge groups could have further charged matter which is not localised at the singularity, which influences the charges. Generally, one obtains one anomaly-free combination generated by
\begin{equation}
Q_{\rm anomaly-free}=\sum_{i}^n \frac{Q_i}{N_i}.
\end{equation}
For our Pati-Salam model on $dP_3$ it turns out by applying equation (\ref{anfree}) that this is the only anomaly-free combination. This $U(1)$ combination will however not turn out to be important and we assume that this cycle is non-trivial in a global embedding and have hence no massless $U(1)$ factor from the abelian symmetries.

%%%%%%%%%%%%%
\section{Majorana masses via D-brane instantons}
\label{sec:NPneutrinos}

We can wrap an $E3$ brane on either the cycle associated with $m_2,$ $m_4$ or $m_{12}$ (cf. Figure~\ref{fig:dp3d7quiver}). We demand that they have the same Chan-Paton factors as the $D3$ branes so that they can be interpreted as gauge instantons. The arguments are similar to the above discussion of $\mu-$term and we directly jump to the discussion of the charged zero modes.

 The cancellation of anomalies does not allow for a solution with all three cycles $m_2,$ $m_4$ or $m_{12}$ wrapped by E3 branes but only a combination of two of them, one example shown in Figure \ref{d3d3d3d37}. However in this example we get a non-perturbative coupling for the remaining neutrino via the higher dimensional non-perturbative coupling $4a-2Ra-2Rb-E3-4a.$ It can be argued that such higher dimensional operators are present \cite{0806.2291}.

 Furthermore to generate a quadratic coupling, such as the Majorana neutrino masses, we have to have two zero modes coupling to $\nu_R,$ which can be achieved by $SP(2)$ instantons \cite{0704.1079,0711.1316}. In addition the unequal ranks ($4$ and $2)$ leading to unequal zero modes for the $E3-D3$ states implies that it is necessary to have $D7-E3$ zero modes to generate a non-vanishing contribution to the superpotential after integrating over the fermionic zero modes. We therefore require states between $E3a,$ $E3b$ and $m_{11}$ $\tilde{\eta}^1,\, \tilde{\eta}^2$ where couplings including $\tilde{\eta}^2$ involve higher dimensional couplings including $\rho_{53}.$ Let us study an example which should be taken just showing the possibility of generating the coupling. The couplings present for the instanton zero modes are shown 
\begin{eqnarray}
&&\eta^1_i Q_{2}^R \bar{\eta}^1_i,\; \eta^1_i \tilde{\eta}^1_i \tilde{\rho}_{73},\; \eta^2_i Q_{3}^R \bar{\eta}^2_i ,\; \eta^2_i \tilde{\eta}^2_i \rho_{53} \tilde{\rho}_{73} ,\; \bar{\eta}^1_i \tilde{\Phi}_{37} \tilde{\bar{\eta}}^1,\; \bar{\eta}^2_i \tilde{\Phi}_{37} \tilde{\bar{\eta}} ^2\, ,
\end{eqnarray}
where $i=1,2.$ To give a contribution proportional to $\nu_R^j\nu_R^j$ we find constraints on the allowed rank for the $D7$ gauge groups. In our notation gauge group $m_{11}$ has to have rank $3$ and gauge group $m_5$ has to have rank $1.$ Other constructions involving $D7$ states coupling to Standard Model fields are dangerous since a vev would induce large unobserved masses for quarks. Furthermore we observe that the vev structure for $D7$ states $\tilde{\Phi}_{37}$ and $\tilde{\rho}_{73}$ requires a breaking of $SU(4)\to SU(3)\times U(1)$ and $SU(2)_R\to U(1):$
\begin{eqnarray}
\tilde{\rho}_{73}&=&\left(\begin{array}{c c c}
\rho & 0 & 0\\
0 & \rho & 0\\
0 & 0 & \rho\\
0& 0 & 0
\end{array}\right), \label{rho7vev}\\
\tilde{\Phi}_{37}&=&\left(\begin{array}{c}
0\\
\phi_{7}
\end{array}\right),\label{phi7vev}
\end{eqnarray}
where $\phi$ and $\rho$ denote some vev which has to be in accordance with the D-flat condition. With this vev structure one can show that after integrating over the charged zero modes according to Equation~\ref{zeromodes} the only quadratic coupling induced in the effective action is the Majorana neutrino mass. We find that all three Majorana masses are distinct due to different $33$ states appearing in the coupling $\rho_{53}$ and $\Phi_{61}.$ Schematically we then have:
\begin{equation}
W_{\rm Majorana}= A(\rho,\phi) e^{-a T_i} \nu_1\nu_1+\frac{\Phi_{61}^2}{\Lambda}A(\rho,\phi)e^{-a T_i} \nu_2\nu_2+\frac{\rho_{53}^2}{\Lambda}A(\rho,\phi) e^{-a T_i} \nu_3\nu_3\, ,
\end{equation}
where $T_i$ denotes the chiral superfield associated to the $4-$cycle volume the $E3$ brane is wrapping. Besides the dependence on $73$ states in $A,$ we expect $A$ not to have any further suppression.

However, for these masses to be present, the vev structure in Equation~\ref{rho7vev} and~\ref{phi7vev} requires that after the breaking to $SU(3)\times U(1)_{B-L}$ and $U(1)_{2R}$ these gauge groups are the diagonal product of the $D3$  and $D7$ gauge groups $U(3)$ or respectively $U(1).$ For the $U(1)$ symmetries to remain massless it is necessary that the abelian $U(1)$ factors in the $D7$ gauge groups remain massless. This is not the case in our setup since the K\"ahler modulus is charged under these $U(1)$ symmetries and hence generates a St\"uckelberg mass for these $U(1)$ symmetries. In this setup it is hence impossible to keep the standard hypercharge in models based on the Pati-Salam gauge groups massless. 

In principle, the hierarchy between the $\mu-$term (electroweak scale) and the Majorana mass (intermediate scale) can arise since they originate from distinct $E3$ branes wrapping different cycles. We shall note that all bi-linear operators in the right-handed quark fields are induced with the same suppression involving different $D3-D7$ fields compared to the right handed neutrino bi-linear. We assume that the $D3-D7$ fields do not obtain a vev and due to the large suppression of that coupling they are phenomenologically irrelevant.

%%%%%%%%%%%%%%%%%%%%%%%%%%%%%%%%%%%%%
\section{Details on gauge coupling unification}
\label{sec:running}
%%%%%%%%%%%%%%%%%%%%%%%%%%%%%%%%%%%%%

In this appendix we give the detailed conventions and expressions used for the gauge coupling running and bother about details on the $U(1)$ normalisation factors. We take the normalisation of the generators of the fundamental representation of $U(N)$ to be ${\rm Tr}{(T^a T^b)}=1/2\delta^{ab}.$ In this convention we have for the $SU(2)$ generators $T^a=\frac{\sigma^a}{2},$ where $\sigma^a$ are the standard Pauli matrices. The diagonal generator in $SU(4)$ $T^{15}$ leading to $U(1)_{B-L}$ is given by
\begin{equation}
T^{15}=\left(\begin{array}{c c c c}\frac{1}{\sqrt{24}}&0 & 0 & 0\\
0 & \frac{1}{\sqrt{24}} & 0 & 0 \\
0 & 0 & \frac{1}{\sqrt{24}} & 0 \\
0 & 0 & 0 & -\sqrt{\frac{3}{8}}\end{array}\right).
\label{t15}
\end{equation}
We have the following Lagrangian for the gauge field strength and a field $\psi$ transforming in the fundamental of $U(N)$
\begin{equation}
{\cal L}=-\frac{1}{4}F^a_{\mu\nu}F^{\mu\nu a} + \psi^\dagger \gamma^\mu \left(\partial_\mu+igA^a_\mu T^a\right)\psi\, .
\end{equation}
%%%%%%%%%%%%%%%%%%%%%%%%%%%%%%%%%%%%%%
\subsection*{$\bf{SU(N)_a\times SU(N)_b\to SU(N)_d}$}
%%%%%%%%%%%%%%%%%%%%%%%%%%%%%%%%%%%%%%
We start with the following Lagrangian for the bosonic components of the gauge fields and fields $\psi_{1,2}$ in the fundamental of $SU(N)_{a,b}$ 
\begin{eqnarray}
{\cal L}=\frac{1}{4}F^a_{\mu\nu}F^{\mu\nu a}- \frac{1}{4}F^b_{\mu\nu}F^{\mu\nu b} + \psi_1^\dagger \gamma^\mu \left(\partial_\mu+ig_a A^a_\mu T^a\right)\psi_1+\psi_2^\dagger \gamma^\mu \left(\partial_\mu+ig_bB^b_\mu T^b\right)\psi_2\, .
\end{eqnarray}
Let us define the new gauge fields after breaking in terms of the old generators
\begin{equation}
\left( \begin{array}{c}
A^a_\mu \\
B^a_\mu \end{array}\right)=\left(\begin{array}{c c}\cos{\theta} & -\sin{\theta}\\
\sin{\theta} & \cos{\theta}
\end{array}\right)\left(\begin{array}{c}L^a_\mu \\
M^a_\mu
\end{array}\right),
\end{equation}
where $L^a_\mu$ denotes the generator of the diagonal gauge group and $M^a_\mu$ the orthogonal massive $SU(N)$ generators. With this definition we see that the Lagrangian becomes
\begin{eqnarray}
\nonumber {\cal L}&=&\frac{1}{4}F^a_{\mu\nu L}F_L^{\mu\nu a}- \frac{1}{4}F^b_{\mu\nu M}F^{\mu\nu b}_M + \psi_1^\dagger \gamma^\mu \left(\partial_\mu+ig_a [\cos{(\theta)}L^a_\mu-\sin{(\theta)}M^a_\mu]\, T^a\right)\psi_1\\
\nonumber &&+\psi_2^\dagger \gamma^\mu \left(\partial_\mu+ig_b[\sin{(\theta)}L^a_\mu+\cos{(\theta)}M^a_\mu]\, T^a\right)\psi_2\\
\nonumber&=&\frac{1}{4}F^a_{\mu\nu L}F_L^{\mu\nu a}- \frac{1}{4}F^b_{\mu\nu M}F^{\mu\nu b}_M + \psi_1^\dagger \gamma^\mu \left(\partial_\mu+ig [L^a_\mu-\tan{(\theta)}M^a_\mu]\, T^a\right)\psi_1\\
&&+\psi_2^\dagger \gamma^\mu \left(\partial_\mu+ig[L^a_\mu+\cot{(\theta)}M^a_\mu]\, T^a\right)\psi_2\, ,
\end{eqnarray}
where we used the definition of the diagonal gauge coupling in terms of the old gauge coupling
\begin{equation}
g_d=g_a \cos{\theta}\, ,\; g_d=g_b \sin{\theta}\, .
\end{equation}
Independent on the breaking scale we find for the inverse gauge couplings the following boundary condition
\begin{equation}
\alpha^{-1}_d=\alpha^{-1}_a+\alpha^{-1}_b\, . \label{boundarydiagonal}
\end{equation}
%%%%%%%%%%%%%%%%%%%%%%%%%%%%%%%%%%%%%%
\subsection*{$\bf{SU(4)\to SU(3)\times U(1)_{B-L}}$}
%%%%%%%%%%%%%%%%%%%%%%%%%%%%%%%%%%%%%%
We start with a Lagrangian for the bosonic gauge field and a fermionic field $\psi$ in the fundamental of $SU(4)$
\begin{equation}
{\cal L}=-\frac{1}{4}F^a_{\mu\nu}F^{\mu\nu a}+\psi^\dagger \gamma^\mu(\partial_\mu+i g_4 A^a_\mu T^a)\psi\, .
\end{equation}
After breakdown to $SU(3)\times U(1)_{B-L}$ the surviving generators are the usual Gell-Mann generators and the diagonal generator $T^{15}$ (cf. Equation~\ref{t15}). $\psi$ decomposes into $q$ transforming as $(3,1/3)$ and $e$ $(1,-1).$ We hence can write after breaking
\begin{eqnarray}
\nonumber{\cal L}&=&-\frac{1}{4}(F_3)^2-\frac{1}{4}(F_{15})^2+\psi^\dagger \gamma^\mu(\partial_\mu+i g_4 A^{\alpha}_\mu \lambda^a+ig_4 A^{15}_\mu T^{15})\psi\\
\nonumber&=&-\frac{1}{4}(F_3)^2-\frac{1}{4}(F_{15})^2+q^\dagger \gamma^\mu(\partial_\mu+i g_4 A^{\alpha}_\mu \lambda^a+ig_4 A^{15}_\mu T^{15})q\\
\nonumber &&+e^\dagger \gamma^\mu(\partial_\mu+i g_4 A^{\alpha}_\mu \lambda^a+ig_4 A^{15}_\mu T^{15})e\\
\nonumber&=&-\frac{1}{4}(F_3)^2-\frac{1}{4}(F_{15})^2+q^\dagger \gamma^\mu(\partial_\mu+i g_4 A^{\alpha}_\mu \lambda^a+\frac{i}{3}g_4 A^{15}_\mu \sqrt{\frac{3}{8}}Q_{B-L})q\\ \nonumber &&+e^\dagger \gamma^\mu(\partial_\mu+i g_4 A^{\alpha}_\mu \lambda^a-ig_4 A^{15}_\mu \sqrt{\frac{3}{8}}Q_{B-L})e\\
\nonumber &=&-\frac{1}{4}(F_3)^2-\frac{1}{4}(F_{15})^2+q^\dagger \gamma^\mu(\partial_\mu+i g_4 A^{\alpha}_\mu \lambda^a+\frac{i}{3}g_{B-L} A^{15}_\mu Q_{B-L})q\\  &&+e^\dagger \gamma^\mu(\partial_\mu+i g_4 A^{\alpha}_\mu \lambda^a-ig_{B-L} A^{15}_\mu Q_{B-L})e\, ,
\end{eqnarray}
where we defined
\begin{equation}
g_{B-L}=\sqrt{\frac{3}{8}}g_4\,.
\end{equation}
This directly leads to
\begin{equation}
\boxed{\alpha^{-1}_{B-L}=\frac{8}{3}\alpha^{-1}_4\, .}
\label{boundbml}
\end{equation}
%%%%%%%%%%%%%%%%%%%%%%%%%%%%%%%%%%%%%%
\subsection*{${\bf SU(2)_R\times U(1)_{B-L}\to U(1)_{Y}}$}
%%%%%%%%%%%%%%%%%%%%%%%%%%%%%%%%%%%%%%
Recall that $U(1)_Y$ is generated by $Q_Y=1/2 Q_{B-L}+T_3^{2R}.$ This leads to the conventional charges with the left-handed quarks having charge $1/6.$ 
Starting with the following Lagrangian
\begin{eqnarray}
\nonumber {\cal L}&=& -\frac{1}{4} (F_{B-L})^2-\frac{1}{4} (F_{2R})^2+\psi_1^\dagger\gamma^\mu(\partial_\mu+ig_{B-L}A_\mu Q_{B-L})\psi_1+\psi_2^\dagger\gamma^\mu(\partial_\mu+ig_{2R}A_\mu^{2R} T^a)\psi_2\\
\nonumber &=& -\frac{1}{4} (F_{B-L})^2-\frac{1}{4} (F_{2R})^2+\psi_1^\dagger\gamma^\mu(\partial_\mu+2ig_{B-L}A_\mu \frac{Q_{B-L}}{2})\psi_1\\ &&+\psi_2^\dagger\gamma^\mu(\partial_\mu+ig_{2R}A_\mu^{2R} T^a)\psi_2
\end{eqnarray}
In analogy to the discussion of the breaking of the diagonal gauge groups we then find $g_Y$ in terms of the previous gauge couplings to be given by
\begin{equation}
2 g_{B-L}\sin{\varphi}=g_y\, , \; g_{2R}\cos{\varphi}=g_y\, .
\end{equation}
This leads to
\begin{equation}
\boxed{\alpha^{-1}_Y=\alpha_{2R}^{-1}+\frac{1}{4} \alpha^{-1}_{B-L}\, .}
\label{boundy}
\end{equation}
%%%%%%%%%%%%%%%%%%%%%%%%%%%%%%%%%%%%%%
\subsection*{${\bf SU(2)_L\times U(1)_Y\to U(1)_{em}}$}
%%%%%%%%%%%%%%%%%%%%%%%%%%%%%%%%%%%%%%
Electromagnetic $U(1)$ then is generated by $Q_{em}=Q_Y+T_3^L.$ In the same fashion as before in the case of breaking two gauge groups to the diagonal gauge group, one finds for the gauge coupling of electromagnetism $e$ in terms of $g_Y$ and $g_{2L}$ as follows
\begin{equation}
e=g_Y \cos{\theta_W}\, ,\; e=g_{2L} \sin{\theta_W}\, .
\end{equation}
This leads to the boundary conditions for the inverse gauge couplings at $M_Z$
\begin{equation}
\boxed{\alpha^{-1}_{em}=\alpha^{-1}_Y+\alpha^{-1}_{2L}\, .}
\end{equation}
%%%%%%%%%%%%%%%%%%%%%%%%%%%%%%%%%%%%%%
\subsection*{Summary of gauge coupling running in $dP_3$}
%%%%%%%%%%%%%%%%%%%%%%%%%%%%%%%%%%%%%%
This is a summary of the running of gauge couplings in the model on $dP_3.$ For the original gauge groups it becomes
\begin{eqnarray}
\alpha^{-1}_{4a,b}(x)&=&\alpha^{-1}(M_s)+\frac{\beta_{4a,b}}{2\pi}\log{\left(\frac{M_s}{x}\right)}\;,\text{ for }M_{4d}\leq x\leq M_{s}\\
\alpha^{-1}_{2La,b}(x)&=&\alpha^{-1}(M_s)+\frac{\beta_{2La,b}}{2\pi}\log{\left(\frac{M_s}{x}\right)}\;,\text{ for }M_{2Ld}\leq x\leq M_{s}\\
\alpha^{-1}_{2Ra,b}(x)&=&\alpha^{-1}(M_s)+\frac{\beta_{2Ra,b}}{2\pi}\log{\left(\frac{M_s}{x}\right)}\;,\text{ for }M_{2Rd}\leq x\leq M_{s}
\end{eqnarray}
We find the following $\beta-$function coefficients
\begin{eqnarray}
\beta_{4a}&=&(-3/2+m_6,-7),\,\beta_{4b}=(-9/2,-7)\,,\,\\
\beta_{2La}&=&(m_6,-1),\,\beta_{2Lb}=(4+m_6-m_9,0)\, ,\\
\beta_{2Ra}&=&(5/2+m_6,0),\,\beta_{2Rb}=(-1/2,-1)\, ,
\end{eqnarray}
where the first number denotes the gauge coupling including $D7$ states in the running and the second one involves no $D7$ states in the running.
After breaking to the diagonal gauge groups we have the following running for the gauge coupling running
\begin{eqnarray}
\nonumber\alpha^{-1}_{4d}(x)&=&\alpha^{-1}(M_{4d})+\frac{\beta_{4d}}{2\pi}\log{\left(\frac{M_{4d}}{x}\right)}\\ &=&2\alpha^{-1}(M_s)+\frac{\beta_{4a}+\beta_{4b}}{2\pi}\log{\left(\frac{M_s}{M_{4d}}\right)}+\frac{\beta_{4d}}{2\pi}\log{\left(\frac{M_{4d}}{x}\right)}\;,\text{ for }M_{4}\leq x\leq M_{4d}\\
\alpha^{-1}_{2L}(x)&=&\alpha^{-1}(M_{2Ld})+\frac{\beta_{2L}}{2\pi}\log{\left(\frac{M_{2Ld}}{x}\right)}\\ \nonumber &=&2\alpha^{-1}(M_s)+\frac{\beta_{2La}+\beta_{2Lb}}{2\pi}\log{\left(\frac{M_s}{M_{2Ld}}\right)}+\frac{\beta_{2Ld}}{2\pi}\log{\left(\frac{M_{2Ld}}{x}\right)}\;,\text{ for }M_{Z}\leq x\leq M_{2Ld} \label{su2lrunning}\\
\alpha^{-1}_{2R}(x)&=&\alpha^{-1}(M_{2Rd})+\frac{\beta_{2R}}{2\pi}\log{\left(\frac{M_{2Rd}}{x}\right)}\\ \nonumber &=&2\alpha^{-1}(M_s)+\frac{\beta_{2Ra}+\beta_{2Rb}}{2\pi}\log{\left(\frac{M_s}{M_{2Rd}}\right)}+\frac{\beta_{2Rd}}{2\pi}\log{\left(\frac{M_{2Rd}}{x}\right)}\;,\text{ for }M_{Z}\leq x\leq M_{2Rd}
\end{eqnarray}
Assuming all $D7$ states being integrated out, we find the following $\beta-$function coefficients
\begin{eqnarray}
\beta_{4d}&=&(-6,-6)\, ,\\
\beta_{2Ld}&=&(3,1)\, ,\\
\beta_{2Rd}&=&(3,1)\, ,
\end{eqnarray}
where the second number corresponds to the $\beta-$function with only one Higgs generation as opposed to three. After breaking of the $SU(4)\to SU(3)\times U(1)_{B-L}$ at $M_4$ we have the following running for the gauge couplings
\begin{eqnarray}
\nonumber\alpha^{-1}_{3}(x)&=&\alpha^{-1}_3(M_{4})+\frac{\beta_3}{2\pi}\log{\left(\frac{M_4}{x}\right)}=2\alpha^{-1}(M_s)+\frac{\beta_{4a}+\beta_{4b}}{2\pi}\log{\left(\frac{M_s}{M_{4d}}\right)}\\ &&+\frac{\beta_{4d}}{2\pi}\log{\left(\frac{M_{4d}}{M_4}\right)}+\frac{\beta_3}{2\pi}\log{\left(\frac{M_4}{x}\right)}\;,\text{ for }M_{Z}\leq x\leq M_{4}\\
\nonumber\alpha^{-1}_{B-L}(x)&=&\alpha^{-1}_{B-L}(M_{4})+\frac{\beta_{B-L}}{2\pi}\log{\left(\frac{M_4}{x}\right)}=\frac{8}{3}\alpha^{-1}_{4}(M_{4})+\frac{\beta_{B-L}}{2\pi}\log{\left(\frac{M_4}{x}\right)}\;,\text{ for }M_{Y}\leq x\leq M_{4}\\
&=&\frac{16}{3}\alpha^{-1}(M_s)+\frac{4(\beta_{4a}+\beta_{4b})}{3\pi}\log{\left(\frac{M_s}{M_{4d}}\right)}+\frac{4\beta_{4d}}{3\pi}\log{\left(\frac{M_{4d}}{M_4}\right)}+\frac{\beta_{B-L}}{2\pi}\log{\left(\frac{M_4}{x}\right)}
\end{eqnarray}
where we used the boundary condition found in Equation~\ref{boundbml} for the running of $U(1)_{B-L}.$ We find the following $\beta-$function coefficients
\begin{equation}
\beta_3=-3\, ,\; \beta_{B-L}=16\, .
\end{equation}
After breaking $SU(2)_R\times U(1)_{B-L}\to U(1)_Y,$ where $Q_Y=T^3_3+Q_{B-L}/2,$ we have the following running for the inverse gauge coupling for $U(1)_Y$
\begin{eqnarray}
\nonumber \alpha^{-1}_Y&=&\alpha^{-1}_{2R}(M_Y)+\frac{1}{4}\alpha^{-1}_{B-L}(M_Y)+\frac{\beta_Y}{2\pi}\log{\left(\frac{M_Y}{x}\right)}\;,\text{ for }M_{Z}\leq x\leq M_{Y}\\
\nonumber &=&\frac{10}{3}\alpha^{-1}(M_s)+\frac{\beta_{2Ra}+\beta_{2Rb}}{2\pi}\log{\left(\frac{M_s}{M_{2Rd}}\right)}+\frac{\beta_{2Rd}}{2\pi}\log{\left(\frac{M_{2Rd}}{M_Y}\right)}+\frac{(\beta_{4a}+\beta_{4b})}{3\pi}\log{\left(\frac{M_s}{M_{4d}}\right)}\\ &&+\frac{\beta_{4d}}{3\pi}\log{\left(\frac{M_{4d}}{M_4}\right)}+\frac{\beta_{B-L}}{8\pi}\log{\left(\frac{M_4}{M_Y}\right)}+\frac{\beta_Y}{2\pi}\log{\left(\frac{M_Y}{x}\right)}\, .
\end{eqnarray}
For a MSSM matter content, the $\beta-$function should be the standard one of $11$ whereas with three Higgs generations it is $13$
\begin{equation}
\beta_Y=(11,13)\, .
\end{equation}
Note that in the case when we break to the MSSM gauge group at the string scale (integrating out the additional Higgs generations) the running reduces precisely to the running of the gauge couplings in the MSSM with the following boundary conditions
\begin{equation}
\alpha^{-1}_3=\alpha^{-1}_2=2\alpha_i^{-1}\; \text{ and } \alpha^{-1}_Y=\frac{10}{3}\alpha_i^{-1}\, .
\end{equation}
%%%%%%%%%%%%%%%%%%%%%%%%%%%%%%%%%%%%%%%
\subsection{Bound on the string scale} \label{sec:couplingestimate}
%%%%%%%%%%%%%%%%%%%%%%%%%%%%%%%%%%%%%%%
To achieve the experimentally observed difference between the $SU(3)$ and $U(1)_Y$ gauge coupling at the weak scale, we find a lower bound on the possible string scale in $dP_3.$ This limit is ought to change when taking threshold and supersymmetry breaking effects into account. However, using the gauge coupling running from the previous sections for now we find
\begin{eqnarray}
\nonumber \alpha^{-1}_Y-\alpha^{-1}_3&=&\frac{10}{3}\alpha^{-1}(M_s)+\frac{\beta_{2Ra}+\beta_{2Rb}}{2\pi}\log{\left(\frac{M_s}{M_{2Rd}}\right)}+\frac{\beta_{2Rd}}{2\pi}\log{\left(\frac{M_{2Rd}}{M_Y}\right)}\\ \nonumber &&+\frac{(\beta_{4a}+\beta_{4b})}{3\pi}\log{\left(\frac{M_s}{M_{4d}}\right)}+\frac{\beta_{4d}}{3\pi}\log{\left(\frac{M_{4d}}{M_4}\right)}+\frac{\beta_{B-L}}{8\pi}\log{\left(\frac{M_4}{M_Y}\right)}+\frac{\beta_Y}{2\pi}\log{\left(\frac{M_Y}{M_Z}\right)}\\ \nonumber &&-
\left(2\alpha^{-1}(M_s)+\frac{\beta_{4a}+\beta_{4b}}{2\pi}\log{\left(\frac{M_s}{M_{4d}}\right)}+\frac{\beta_{4d}}{2\pi}\log{\left(\frac{M_{4d}}{M_4}\right)}+\frac{\beta_3}{2\pi}\log{\left(\frac{M_4}{M_Z}\right)}\right)
\\
\nonumber &=&\frac{4}{3}\alpha^{-1}(M_s)+\frac{\beta_{2Ra}+\beta_{2Rb}}{2\pi}\log{\left(\frac{M_s}{M_{2Rd}}\right)}+\frac{\beta_{2Rd}}{2\pi}\log{\left(\frac{M_{2Rd}}{M_Y}\right)}-\frac{(\beta_{4a}+\beta_{4b})}{6\pi}\log{\left(\frac{M_s}{M_{4d}}\right)}\\
&& -\frac{\beta_{4d}}{6\pi}\log{\left(\frac{M_{4d}}{M_4}\right)}+\frac{\beta_{B-L}}{8\pi}\log{\left(\frac{M_4}{M_Y}\right)}+\frac{\beta_Y}{2\pi}\log{\left(\frac{M_Y}{M_Z}\right)}-\frac{\beta_3}{2\pi}\log{\left(\frac{M_4}{M_Z}\right)}
\label{diff1}
\end{eqnarray}
 We now distinguish the following cases, involving first running with $D3-D7$ brane states and in the second without. In both cases the difference is larger when we include all three Higgs generations.
 \subsubsection*{Estimate with $D3-D7$ states}
 To maximise the difference we find the following two cases depending on the values for $m_{6}$
 \begin{eqnarray}
\text{Case A: }  \beta_{2Ra}+\beta_{2Rb}\geq \beta_{2Rd} , & & (2+m_6\geq 3) \Rightarrow M_{2Rd}=M_Y\\
\text{Case B: } \beta_{2Ra}+\beta_{2Rb}\leq \beta_{2Rd} , & & (2+m_6\leq 3) \Rightarrow M_{2Rd}=M_s\\
 \beta_{4a}+\beta_{4b}\geq \beta_{4d}, & & (-5+m_6\geq -6) \Rightarrow M_{4d}=M_s
 \end{eqnarray}
 In the first limit Equation~\ref{diff1} can be rewritten as
 \begin{eqnarray}
\nonumber \alpha^{-1}_Y-\alpha^{-1}_3&=&\frac{4}{3}\alpha^{-1}(M_s)+\log{(M_s)}\left(\frac{4+m_6}{2\pi}\right)+\log{(M_4)}\left(\frac{5}{2\pi}\right)\\ \nonumber &&+\log{(M_Y)}\left(\frac{7-m_6}{2\pi}\right)-\log{(M_Z)}\left(\frac{8}{\pi}\right)\\
 &\leq&\frac{4}{3}\alpha^{-1}(M_s)+\log{\left(\frac{M_s}{M_Z}\right)}\left(\frac{8}{\pi}\right)
 \end{eqnarray}
 where in the last line we maximised the difference by taking the breaking scales $M_{4}$ and $M_{Y}$ to their maximal value $M_{s}.$ Similarly in the second limit we observe
  \begin{eqnarray}
\nonumber \alpha^{-1}_Y-\alpha^{-1}_3&=&\frac{4}{3}\alpha^{-1}(M_s)+\log{(M_s)}\left(\frac{5}{2\pi}\right)+\log{(M_4)}\left(\frac{5}{2\pi}\right)\\ \nonumber &&+\log{(M_Y)}\left(\frac{6}{2\pi}\right)-\log{(M_Z)}\left(\frac{8}{\pi}\right)\\
 &\leq&\frac{4}{3}\alpha^{-1}(M_s)+\log{\left(\frac{M_s}{M_Z}\right)}\left(\frac{8}{\pi}\right)
 \end{eqnarray}
 where the upper limit appears again in the limit $M_{Y}$ and $M_{4}$ approaching the string scale $M_{s}.$
 %%%%%%%%%%%%%%%%%%%%%%%%%%%%%%%%%%%%%%
  \subsubsection*{Estimate without $D3-D7$ states}
 %%%%%%%%%%%%%%%%%%%%%%%%%%%%%%%%%%%%%%
  In this case we observe
  \begin{eqnarray}
 \beta_{2Ra}+\beta_{2Rb}\geq \beta_{2Rd}, & & (-1< 3) \Rightarrow M_{2Rd}=M_s\\
 \beta_{4a}+\beta_{4b}< \beta_{4d}, & & (-14< -6) \Rightarrow M_{4d}=M_4
 \end{eqnarray}
 In this limit Equation~\ref{diff1} can be rewritten as
 \begin{eqnarray}
\nonumber \alpha^{-1}_Y-\alpha^{-1}_3&=&\frac{4}{3}\alpha^{-1}(M_s)+\log{(M_s)}\left(\frac{9+14}{6\pi}\right)+\log{(M_4)}\left(\frac{7}{6\pi}\right)\\ \nonumber &&+\log{(M_Y)}\left(\frac{6}{2\pi}\right)-\log{(M_Z)}\left(\frac{8}{\pi}\right)\\
 &\leq&\frac{4}{3}\alpha^{-1}(M_s)+\log{\left(\frac{M_s}{M_Z}\right)}\left(\frac{8}{\pi}\right)
 \end{eqnarray}

 In consistency with the breaking pattern, we find in both cases the same lower bound on the string scale.
 
 %%%%%%%%%%%%%%%%%%%%%%%%%%%%%%%%%%%%%%
 \subsubsection*{Estimate of $\alpha^{-1}(M_{s})$}
 %%%%%%%%%%%%%%%%%%%%%%%%%%%%%%%%%%%%%%
The above estimate contains a term proportional to $\alpha^{-1}(M_{s})$ and having in mind that we are able to tune $\alpha^{-1}(M_{s})$ it is interesting to estimate its size given the experimental input of the gauge couplings at $M_{Z}.$ One straight forward constraint can be obtained from the gauge coupling of $SU(2)_{L}$ as given in Equation~\ref{su2lrunning}
\begin{eqnarray}
\nonumber 29.57=\alpha_{2L}^{-1}(M_{Z})&=&2\alpha^{-1}(M_s)+\frac{\beta_{2La}+\beta_{2Lb}}{2\pi}\log{\left(\frac{M_s}{M_{2Ld}}\right)}+\frac{\beta_{2Ld}}{2\pi}\log{\left(\frac{M_{2Ld}}{M_{Z}}\right)}\\
&\geq& 2\alpha^{-1}(M_s)-\frac{1}{2\pi}\log{\left(\frac{M_{s}}{M_{Z}}\right)}\, ,
\end{eqnarray}
where in the last step our choice of beta functions is such that it would give the largest negative contribution due to the running, hence allowing the maximal value for $\alpha^{-1}(M_s).$ We hence find the following upper bound on $\alpha^{-1}(M_{s})$
\begin{equation}
\alpha^{-1}(M_{s})\leq 17.36\, .
\end{equation}
This upper bound is achieved in the limit $M_{2Ld}\to M_{Z}$ and no $D7$ states in the running.
Furthermore this estimate now enables us to give a lower bound on the possible string respectively breaking scales by combining it with the above estimate as discussed in the main text. 

%%%%%%%%%%%%%%%%%%%%%%%%%%%%%%%%%%%%%
\section{Pati-Salam without $D7$ branes}
\label{sec:nod7s}
%%%%%%%%%%%%%%%%%%%%%%%%%%%%%%%%%%%%%
The model discussed in the main part of the text requires a large $D7$ sector due to anomaly cancellation. One might wonder whether a different choice of $D3$ gauge groups can evade this bound. Figure~\ref{ultralocaldp3} shows a $U(4)^4\times U(2)^2$ gauge theory on $dP_3$ that satisfies the anomaly cancellation condition with unequal gauge group ranks. This is an extension of the Pati-Salam model in the main section in the left-right $U(2)$ factors. For a realistic model the breakdown to the Standard Model is crucial and it turns out, using the techniques available to date, that it requires the re-introduction of $D7$ branes. We hence at this stage do not pursue this option any further.
\begin{center}
\includegraphics[width=0.5\textwidth]{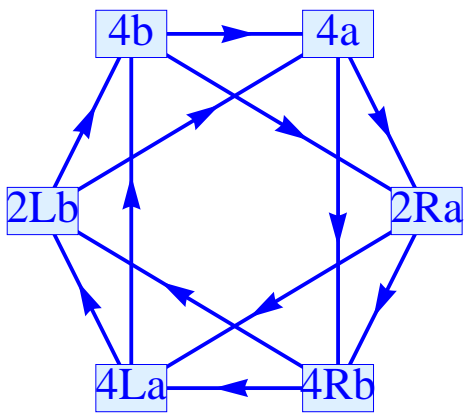}
\captionof{figure}{\footnotesize{A Pati-Salam model on $dP_3$ which does not require $D7$ branes from anomaly cancellation.}\label{ultralocaldp3}}
\end{center}

\bibliographystyle{JHEP}
\bibliography{delPezzo4}
\end{document}